\documentclass[a4paper,11pt]{article}
\pdfoutput=1 

\usepackage{jheppub} 

\usepackage[T1]{fontenc} 

\newcommand{\pip}{\ensuremath{\pi^{+}}}
\newcommand{\pim}{\ensuremath{\pi^{-}}}
\newcommand{\pipm}{\ensuremath{\pi^{\pm}}}
\newcommand{\pimp}{\ensuremath{\pi^{\mp}}}
\newcommand{\piz}{\ensuremath{\pi^{0}}}
\newcommand{\rhoz}{\ensuremath{\rho^{0}}}
\newcommand{\aone}{\ensuremath{a_{1}^{\pm}}}

\newcommand{\Bp}{\ensuremath{B^{+}}}
\newcommand{\Bz}{\ensuremath{B^{0}}}
\newcommand{\Bzb}{\ensuremath{\bar B^{0}}}

\newcommand{\Acp}{\ensuremath{{\cal A}_{CP}}}
\newcommand{\Ccp}{\ensuremath{{\cal C}_{CP}}}
\newcommand{\Scp}{\ensuremath{{\cal S}_{CP}}}
\newcommand{\phitwo}{\ensuremath{\phi_{2}}}

\title{\boldmath Resolving the \phitwo\ ($\alpha$) ambiguity in $B \to \rho \rho$}

\author{J. Dalseno}
\affiliation{Instituto Galego de F\'{i}sica de Altas Enerx\'{i}as (IGFAE), Universidade de Santiago de Compostela,\\R\'{u}a de Xoaqu\'{i}n D\'{i}az de R\'{a}bago, Santiago de Compostela, Spain}

\collaborationImg{\hspace{-29pt}\includegraphics[height=100pt,width=!]{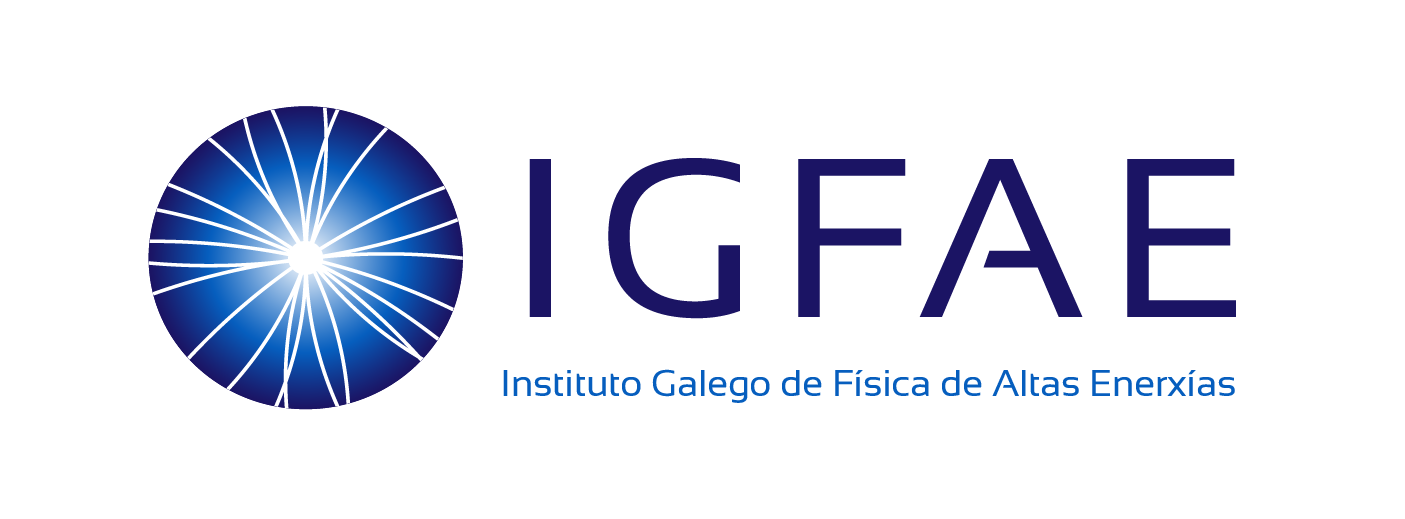}\vspace{15pt}}

\emailAdd{jeremy.peter.dalseno@cern.ch}

\abstract{
  I propose an alternative method for measuring the $CP$ violating phase $\phi_2$ ($\alpha$) without ambiguity in an extended SU(2) isospin triangle analysis, which can ultimately be achieved by exploiting interference effects between $B^0 \to \rho^0 \rho^0$ and $B^0 \to a_1^\pm \pi^\mp$ in a time-dependent flavour-tagged amplitude analysis. Under certain assumptions on the effective $\phi_2$ in each channel, I demonstrate with an idealised amplitude model that potential deviations in the measured $\phi_2$ due to penguin contamination in $B^0 \to a_1^\pm \pi^\mp$ are sufficiently large within current experimental uncertainties that this programme could be executed with Run 3 data at LHCb and easily at Belle II.
}

\keywords{CKM angle alpha, e+-e- Experiments}
\arxivnumber{1808.09391}

\begin{document}
\maketitle
\flushbottom

\section{Introduction}
Violation of the combined charge-parity symmetry ($CP$ violation) in the Standard Model (SM) arises from a single irreducible phase in the Cabibbo-Kobayashi-Maskawa~(CKM) quark-mixing matrix~\cite{Cabibbo,KM}. Various processes offer different yet complementary insight into this phase which manifests in a number of experimental observables over-constraining the Unitarity Triangle. The measurement of such parameters and their subsequent combination is important as New Physics (NP) contributions can present themselves as an inconsistency within the triangle paradigm.

\begin{figure}[h]
  \centering
  \includegraphics[height=115pt,width=!]{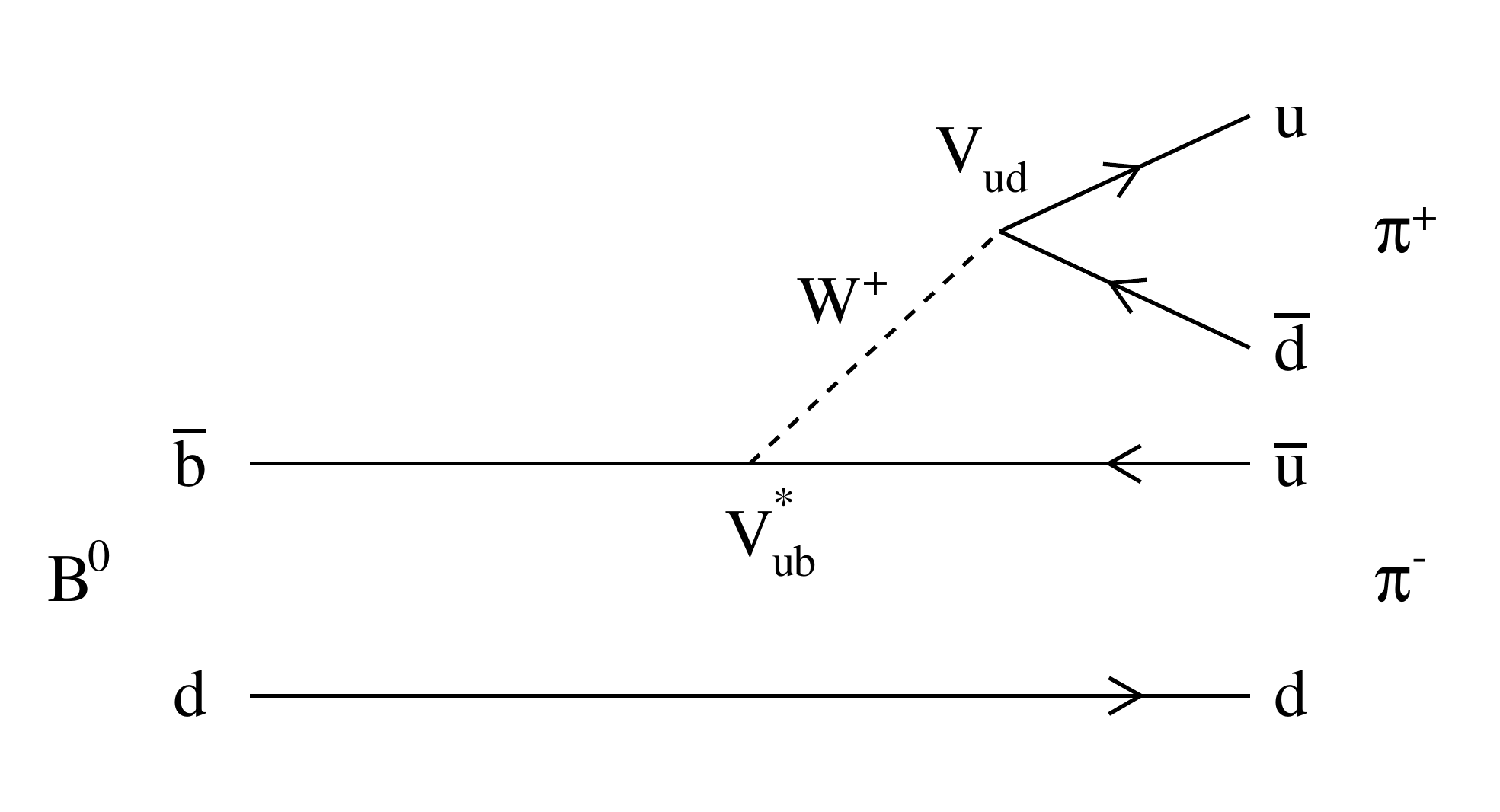}
  \includegraphics[height=115pt,width=!]{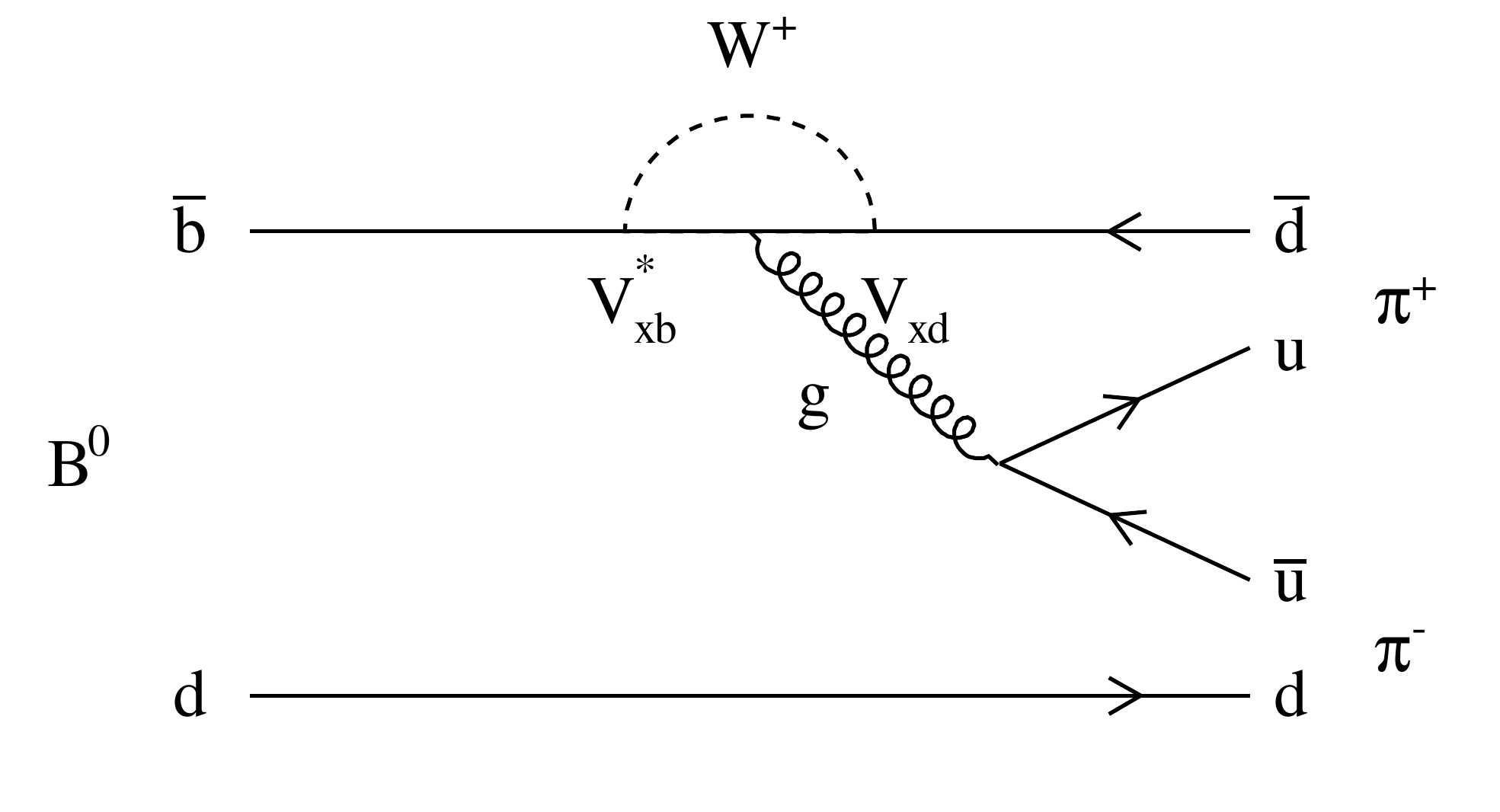}
  \put(-425,110){(a)}
  \put(-210,110){(b)}
  \caption{\label{fig:pipi} Leading-order Feynman diagrams shown producing $\Bz \to \pip \pim$ decays, though the same quark transition can also produce $\Bz \to \rho^{\pm} \pimp$, $\rho^+ \rho^-$ and $\aone \pimp$. (a) depicts the dominant first-order (tree) diagram while (b) shows the second-order loop (penguin) diagram. In the penguin diagram, the subscript $x$ in $V_{xb}$ refers to the flavour of the intermediate-state quark $(x=u,c,t)$.}
\end{figure}

Decays that proceed predominantly through the $\bar b \rightarrow \bar u u \bar d$ tree transition (figure~\ref{fig:pipi}a) in the presence of \Bz--\Bzb\ mixing are sensitive to the interior angle of the Unitarity Triangle $\phitwo = \alpha \equiv \arg(-V_{td}V^{*}_{tb})/(V_{ud}V^{*}_{ub})$, which can be accessed through 
mixing-induced $CP$ violation observables measured from time-dependent, flavour-tagged analyses.
This quark process manifests itself in multiple systems, including $B \to \pi \pi$~\cite{phi2_pipi1,phi2_pipi2,phi2_pipi3,phi2_pipi4,phi2_pipi5,phi2_pipi6}, $\rho \pi$~\cite{phi2_rhopi1,phi2_rhopi2}, $\rho \rho$~\cite{phi2_rhorho1,phi2_rhorho2,phi2_rhorho3,phi2_rhorho4,phi2_rhorho5,phi2_rhorho6,phi2_rhorho7} and $\aone \pimp$~\cite{phi2_a1pi1,phi2_a1pi2,phi2_a1pi3}, where the angle \phitwo\ has so far been constrained with an overall uncertainty of around $4^\circ$~\cite{phi2_gronau,CKMFitter,UTfit}. However, one of the salient features of the overall \phitwo\ combination is the persistence of degenerate solutions within the range $[0, \pi]$: up to the $2\sigma$ level, two solutions currently remain, while beyond this further solutions emerge.

In this paper, I propose a method to resolve the \phitwo\ solution degeneracy in the $B \to \rho \rho$ system, by harnessing interference effects unique to multibody decays. I open in section~\ref{sec:su2}, with a description of the SU(2)-based approaches for controlling distortions in experimental \phitwo\ measurements arising from the accompanying strong-penguin processes. This is followed by a discussion on time-dependent amplitude analysis in section~\ref{sec:tdp}, which is the technique ultimately responsible for eliminating multiple solutions. In section~\ref{sec:iso}, I outline the extension to the SU(2) isospin triangle analysis, which would allow a single solution for \phitwo\ to be obtained that is also free of contamination from strong penguins. To demonstrate the capabilities of this proposed concept, section~\ref{sec:model} describes the rudimentary model used to generate pseudo-experiments interfering $\Bz \to \rhoz \rhoz$ against $\Bz \to \aone \pimp$. The results of the pseudo-experiment study are given for various experimental milestones in section~\ref{sec:results} and finally conclusions are drawn in section~\ref{sec:conc}.

\section{\boldmath Strong-penguin containment in \phitwo\ constraints}
\label{sec:su2}

In general, the extraction of \phitwo\ is complicated by the presence of interfering amplitudes that distort the experimentally determined value of \phitwo\ from its SM expectation and would mask any NP phase if not accounted for. These effects primarily include $\bar b \rightarrow \bar d u \bar u$ strong-loop decays (figure~\ref{fig:pipi}b), although isospin-violating processes such as electroweak penguins, $\piz$--$\eta$--$\eta^\prime$ mixing, $\rho^0$--$\omega$--$\phi$ mixing and the finite $\rho$ width~\cite{rhowidth} can also play a role.

It is possible to remove the isospin-conserving component of this contamination by invoking SU(2) arguments. The original method considers the three possible charge configurations of $B \rightarrow \pi\pi$ decays~\cite{pipi_th1}. In the case of strong penguins, a total isospin of $I=1$ is not allowed by Bose-Einstein statistics, leaving only the possibility of $I=0$ as the mediating gluon is an isoscalar. However, in the specific case of $\Bp \rightarrow \pip \piz$, the further limiting projection $I_{3} = 1$ additionally rules out $I=0$, thereby forbidding strong penguin contributions to this channel.

\begin{figure}[tbp]
  \centering
  \includegraphics[height=110pt,width=!]{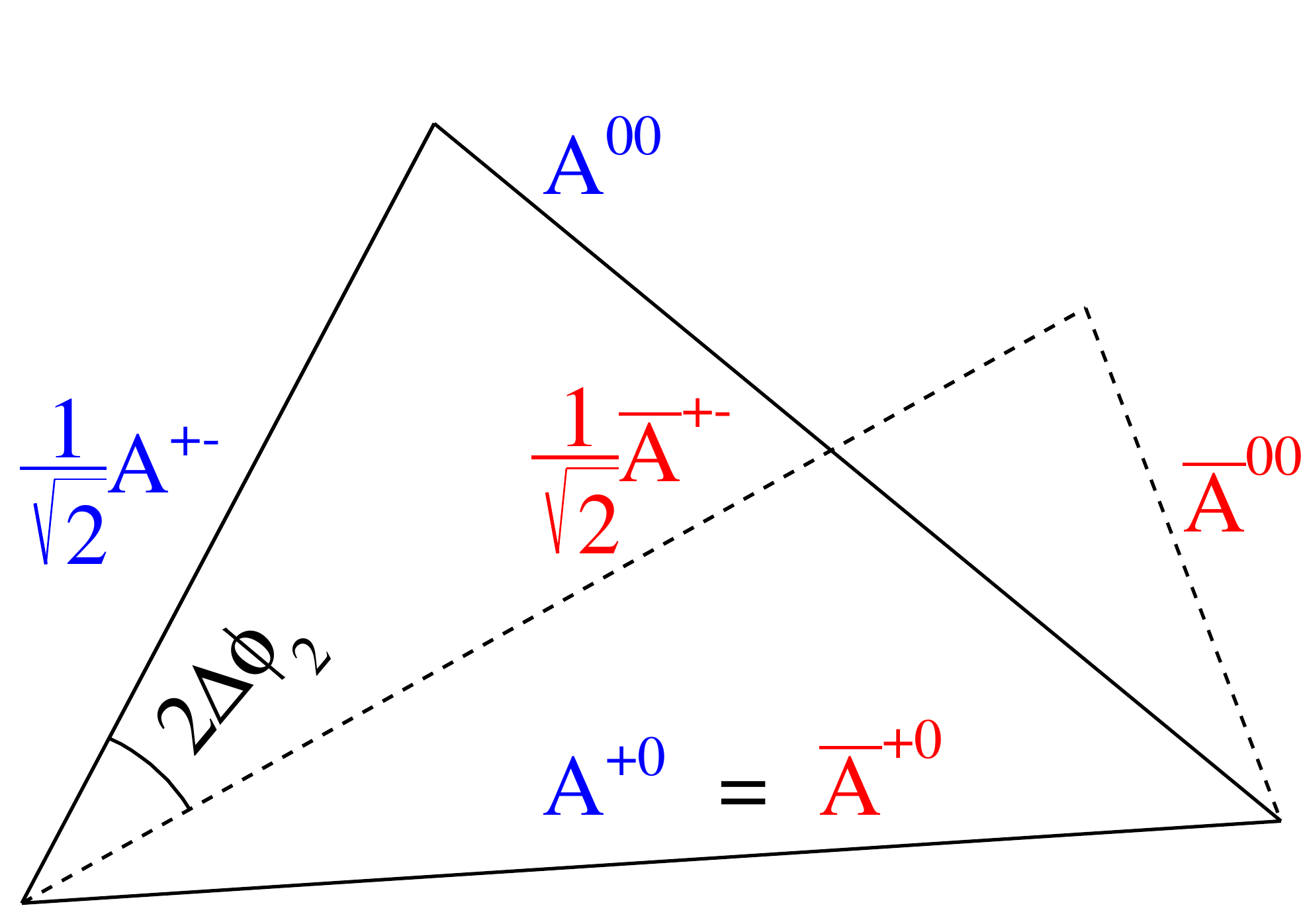}
  \caption{\label{fig_iso} Complex isospin triangles from which $\Delta \phitwo$ can be determined.}
\end{figure}

The complex $B \rightarrow \pi\pi$ and $\bar B \rightarrow \pi\pi$ decay amplitudes obey the isospin relations
\begin{equation}
  A^{+0} = \frac{1}{\sqrt{2}}A^{+-} + A^{00}, \;\;\;\; \bar{A}^{+0} = \frac{1}{\sqrt{2}}\bar{A}^{+-} + \bar{A}^{00},
  \label{eq_iso}
\end{equation}
respectively, where the superscripts refer to the combination of pion charges. The decay amplitudes can be represented as triangles in the complex plane as shown in figure~\ref{fig_iso}. As $\Bp \rightarrow \pip \piz$ is a pure tree mode, these triangles share the same base, $A^{+0}=\bar{A}^{+0}$, allowing the shift in \phitwo\ caused by strong penguin contributions $\Delta \phitwo$, to be determined from the phase difference between $\bar{A}^{+-}$ and $A^{+-}$. These triangles and \phitwo\ can be constrained from the experimentally measured branching fractions, ${\cal B}(\Bz \rightarrow \pip \pim)$, ${\cal B}(\Bz \rightarrow \piz\piz)$ and ${\cal B}(\Bp \rightarrow \pip \piz)$, direct $CP$ violation parameters $\Acp(\Bz \rightarrow \pip \pim)$ and $\Acp(\Bz \rightarrow \piz\piz)$, and the mixing-induced $CP$ violation parameter $\Scp(\Bz \rightarrow \pip\pim)$. This method has an eightfold discrete ambiguity in the determination of \phitwo, which arises from the four triangle orientations about $A^{+0}$ and the two solutions of \phitwo\ through its effective measurement in $\Scp = \sqrt{1 - \Acp^2}\sin (2 \phitwo + 2 \Delta \phitwo)$. In principle, the isospin triangle reflections can be resolved given a favourable measurement of $\Scp(\Bz \rightarrow \piz\piz)$, leaving two overall solutions remaining. However, this would require an experimentally challenging vertex determination from the photons in the final state that have undergone pair production early enough in the detector material~\cite{pipi_Scp_th}.

It was also pointed out soon afterwards that the ambiguity could be resolved in the $\Bz \to (\rho \pi)^0$ system alone without the need to involve the charged $B$ modes~\cite{rhopi_th}, though of course this still can be done in an isospin pentagonal analysis if desired. By performing a time-dependent flavour-tagged amplitude analysis, the additional degrees of freedom afforded by the phase space allow the penguin contribution to be disentangled from the tree amplitudes through interference effects in the Dalitz Plot, exposing a single solution for \phitwo. However at this point, the first-generation $B$ factories have encountered some difficulty obtaining a meaningful constraint with this method due to statistical limitations. Specifically, an ensemble test based on the result obtained by BaBar gives rise to two further solutions that are not statistically distinguishable from their nominal \phitwo~\cite{phi2_rhopi1}, while at Belle, it is not possible to provide uncertainties on \phitwo\ beyond the $1\sigma$ level due to the large experimental uncertainties present~\cite{phi2_rhopi2}. Therefore, it is worth investigating other possibilities to resolve the \phitwo\ ambiguity.

The $B \to \rho \rho$ system has also been studied with the approach created for $B \to \pi\pi$ and is found to dominate the precision on \phitwo\ with only a two-fold degeneracy remaining due to the small penguin contribution which collapses the isospin triangles. As such, I suggest a deeper investigation of the $4\pi$ final state. It has already been pointed out that while the charged and neutral $a_1$ resonances do not overlap in the phase space, a $\Bz \to (\rho\pi)^0$-style analysis for $\Bz \to (a_1\pi)^0 \to \pip \pim \piz \piz$ is still a viable approach through interference with the dominantly longitudinal amplitude of $\Bz \to \rho^+ \rho^-$, though experimentally unattractive due to the presence of two neutral pions in the final state~\cite{a1pi_th1}. However, if a time-dependent flavour-tagged amplitude analysis of $\Bz \to \rhoz \rhoz$ could be performed instead of aiming for its $\Scp$ in the quasi-two-body approximation~\cite{phi2_rhorho5}, the single solution obtained for the effective \phitwo\ could be sufficient to resolve the \phitwo\ ambiguity in the $B \to \rho \rho$ system with the extension to the isospin triangle analysis idea proposed in this paper.

\section{\boldmath Resolving solutions with time-dependent amplitude analysis}
\label{sec:tdp}

The time-dependent decay rates of \Bz\ and \Bzb\ decays to a self-conjugate final state are given by
\begin{eqnarray}
  \label{eq:tdrate}
  \Gamma(t) &\propto& e^{-t/\tau}[(|A|^2 + |\bar A|^2) + (|A|^2 - |\bar A|^2)\cos \Delta m_d t - 2 \Im (\bar A A^*) \sin \Delta m_d t], \nonumber\\
  \bar \Gamma(t) &\propto& e^{-t/\tau}[(|A|^2 + |\bar A|^2) - (|A|^2 - |\bar A|^2)\cos \Delta m_d t + 2 \Im (\bar A A^*) \sin \Delta m_d t],
\end{eqnarray}
respectively, where $A$ is the static decay amplitude, $\tau$ is the \Bz\ lifetime and $\Delta m_d$ is the mass difference between the $B_H$ and $B_L$ mass eigenstates. This form assumes no $CP$ violation in the mixing $|q/p| = 1$, and that the total decay rate difference between the two mass eigenstates is negligible.

In the typical isobar approach, the total amplitude $A$, can be written as the coherent sum over the number of intermediate states in the model with amplitude $A_i$, as a function of 4-body phase space $\Phi_4$,
\begin{equation}
  \label{eq:a}
  A \equiv \sum_i a_iA_i(\Phi_4),
\end{equation}
where $a_i$ is a strong complex coupling  determined directly from the data. Incorporating a complex $CP$ violation parameter $\lambda_i$, for each contribution in the phase space, the total $\bar A$ can be written as
\begin{equation}
  \label{eq:abar}
  \bar A \equiv \sum_i a_i \lambda_i \bar A_i(\Phi_4) = \sum_i a_i \lambda_i A_i(\bar \Phi_4),
\end{equation}
where the phase space has been transformed under $C$ and $P$ conjugation, since $A_i$ should only contain strong dynamics blind to flavour.

As $\Bz \to \rhoz \rhoz$ is an all-pion final state, only intermediate decay channels carrying the weak phase \phitwo, may populate the surrounding phase space in the limit of non-existent penguin contributions. In such a scenario, it is clear that $\lambda = \exp(2i\phitwo)$ can factorise out of the coherent sum given in eq.~\ref{eq:abar}, as it becomes common to all channels. Furthermore, as all intermediate states must be flavour-non-specific, what remains in the conjugate amplitude sum must be equal to the amplitude sum given in eq.~\ref{eq:a} for all points in the phase space, $\Phi_4$.  Then by the final term of eq.~\ref{eq:tdrate}, $\Im(\bar A A^*) = \Im(\lambda A A^*) = \Im(\lambda|A|^2)$. As $|A|^2$ must be real-defined, the imaginary part of the aforementioned product evaluates to $\sin 2\phitwo$, leaving two solutions remaining for \phitwo.

Perhaps ironically, sizeable penguin contamination in the four-charged-pion final state is the key to extracting a single solution for the effective \phitwo\ in $\Bz \to \rhoz \rhoz$ with the time-dependent amplitude analysis technique. Consider that current analyses have yet to find evidence for other significant contributions in the $\Bz \to \rhoz \rhoz$ region, limiting interference opportunities. While there is little doubt these should emerge with increased statistics, it is difficult to envision that the penguin contributions to such amplitudes already suppressed relative to $\Bz \to \rhoz \rhoz$ can be so large as to prevent the weak phase from factorising. However, by expanding the analysis phase space to include $\Bz \to \aone \pimp$, I will demonstrate that the current experimental precision on its physics parameters provides a meaningful space for distortion of its weak phase and direct $CP$ violation parameters from the penguin-free expectations of \phitwo\ and null, respectively. These penguin-induced deviations in turn, allow for a unique determination of the effective \phitwo\ in $\Bz \to \rhoz \rhoz$ through interference, with the amount of data expected to be collected by $B$ physics experiments in the near future.

It should be reminded at this point that isospin-breaking ($I=1$) $\rho$-width effects seem to be best controlled by reducing the $\rho$ analysis window of $\Bz \to \rho^+ \rho^-$ and $\Bp \to \rho^+ \rho^0$ according to the method outlined in ref.~\cite{pipi_th2}. Therefore, I am not advocating for a similar expansion of the analysis phase space in these channels. However, for the specific case of $\Bz \to \rhoz \rhoz$ which cannot be in an $I=1$ state, a larger analysis region for this particular decay would bring the added benefit of possibly removing degenerate \phitwo\ solutions from the overall $B \to \rho\rho$ system.

\section{Extension to the isospin triangle analysis}
\label{sec:iso}

I follow the general approach outlined by the CKMfitter Group given in ref.~\cite{CKMFitter}. They describe 7 mostly independent observables which are related to the decay amplitudes as
\begin{equation}
  \frac{1}{\tau_B^{i+j}} {\cal B}^{ij} = \frac{|\bar A^{ij}|^2 + |A^{ij}|^2}{2}, \hspace{10pt} \Acp^{ij} = \frac{|\bar A^{ij}|^2 - |A^{ij}|^2}{|\bar A^{ij}|^2 + |A^{ij}|^2}, \hspace{10pt} \Scp^{ij} = \frac{2\Im(\bar A^{ij} A^{ij*})}{|\bar A^{ij}|^2 + |A^{ij}|^2},
\end{equation}
where $i,j$ represents the charge configuration of the intermediate $\rho$ resonances and $\tau_B^{i+j}$ is the lifetime of the \Bp\ ($i+j=1$) or \Bz\ ($i+j=0$). For convenience, I also adopt the isospin representation of the amplitude system for function minimisation purposes as given in ref.~\cite{pipi_isorep},
\begin{equation*}
  A^{+0} = \mu e^{i(\Delta-\phitwo)}, \hspace{20pt} \bar A^{+0} = \mu e^{i(\Delta+\phitwo)},
\end{equation*}
\begin{equation*}
  A^{+-} = \mu a, \hspace{20pt} \bar A^{+-} = \mu \bar ae^{2i\phitwo^{+-}},
\end{equation*}
\begin{equation}
  A^{00} = A^{+0} - \frac{A^{+-}}{\sqrt{2}}, \hspace{20pt} \bar A^{00} = \bar A^{+0} - \frac{\bar A^{+-}}{\sqrt{2}},
\end{equation}
where $a$, $\bar a$ and $\mu$ are real, positive parameters related to the magnitude of the decay amplitudes, while $\Delta$ is a relative strong phase. The weak phase $\phitwo^{+-} = \arg(\bar A^{+-}A^{+-*})/2 = \phitwo + \Delta\phitwo$, embodies the shift in \phitwo\ caused by the penguin contamination. The experimental observables can then be related to this set of free parameters as
\begin{equation*}
  {\cal B}^{+0} = \tau_{\Bp} \mu^2, \hspace{20pt} {\cal B}^{+-} = \tau_{\Bz} \mu^2 \frac{a^2 + \bar a^2}{2},
\end{equation*}
\begin{equation*}
  {\cal B}^{00} = \tau_{\Bz} \frac{\mu^2}{4}\biggl\{4 + a^2 + \bar a^2 - 2 \sqrt{2}[a\cos(\phitwo - \Delta) + \bar a \cos (\phitwo + \Delta - 2 \phitwo^{+-})] \biggr\},
\end{equation*}
\begin{equation}
  \Acp^{+-} = \frac{\bar a^2 - a^2}{\bar a^2 + a^2}, \hspace{20pt} \Scp^{+-} = \frac{2 a \bar a}{\bar a^2 + a^2} \sin (2 \phitwo^{+-}).
\end{equation}

Now I extend the method for $\Bz \to \rhoz \rhoz$ by replacing the amplitude-squared-level $\Acp^{00}$ and $\Scp^{00}$ observables that would ordinarily be determined experimentally, with the proposed directly measured amplitude-level observables
\begin{equation}
  |\lambda_{CP}^{00}| = \biggl|\frac{\bar A^{00}}{A^{00}}\biggr|, \hspace{20pt} \phitwo^{00} = \frac{\arg(\bar A^{00} A^{00*})}{2}.
\end{equation}
Clearly, in the limit of vanishing penguin contributions and other isospin-breaking effects, $\phitwo^{+-} = \phitwo^{00} = \phitwo$.

These constructs now permit the isospin triangles and \phitwo\ to be constrained without ambiguity in principle. In this paper, I employ a Frequentist approach where a $\chi^2$ is constructed comparing the above theoretical forms for the 7 observables with their experimentally measured counterparts. The $\Delta \chi^2$ across the range of \phitwo\ can then be converted into a $p$-value scan, assuming it is distributed with one degree of freedom, from which confidence intervals can be derived.

\begin{figure}[tbp]
  \centering
  \includegraphics[height=120pt,width=!]{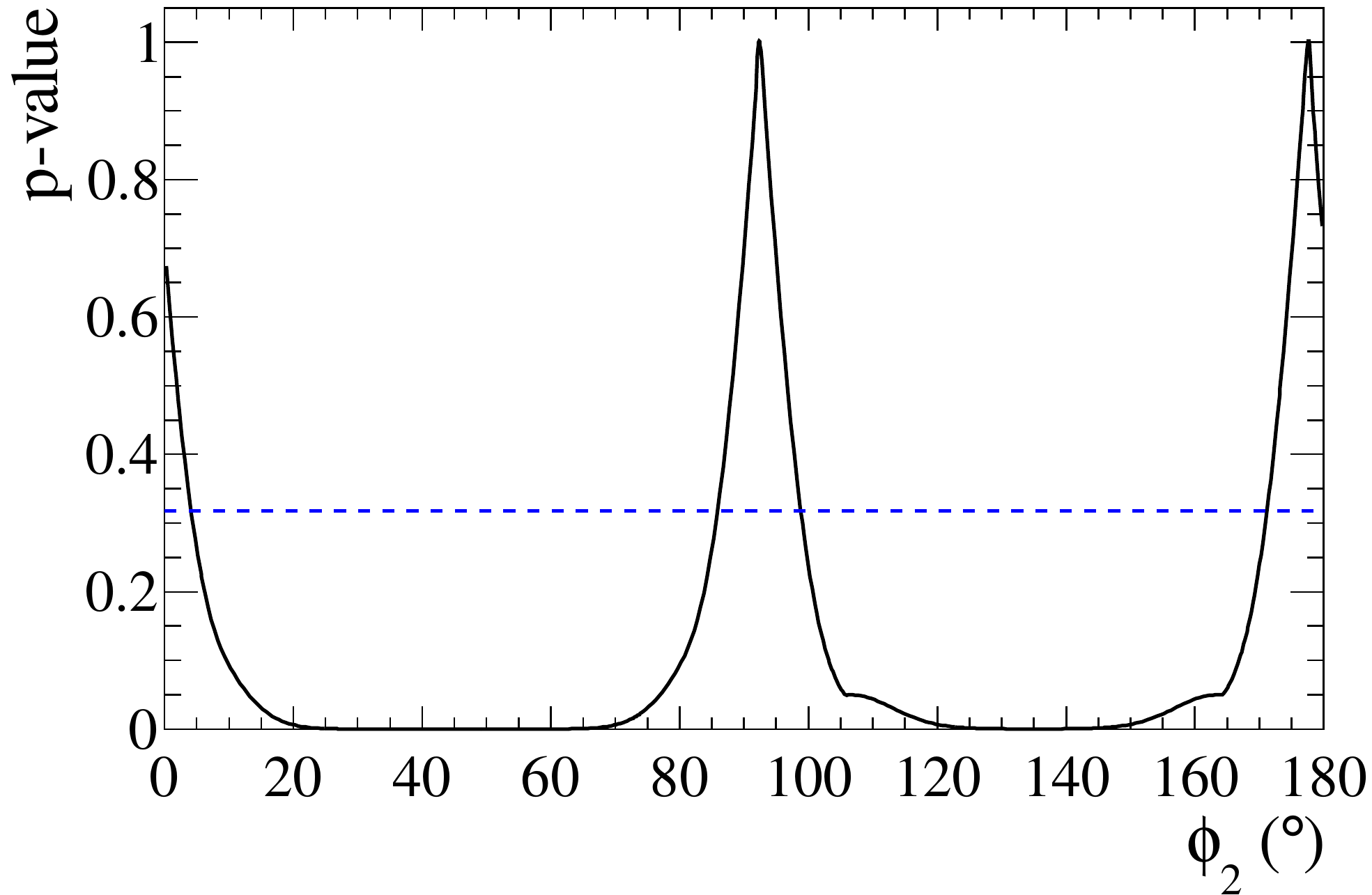}
  \includegraphics[height=120pt,width=!]{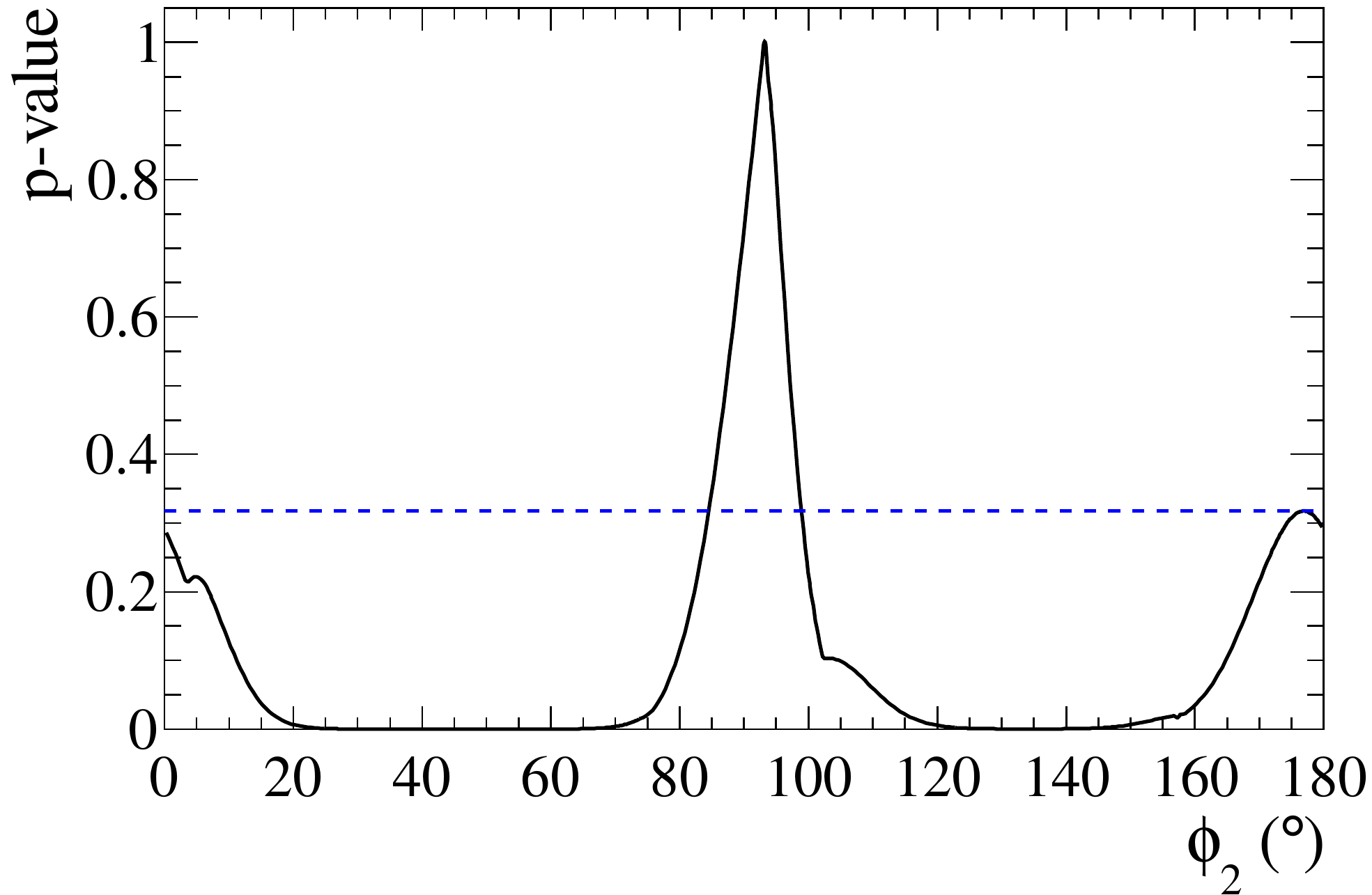}
  \put(-215,105){(a)}
  \put(-28,105){(b)}

  \includegraphics[height=120pt,width=!]{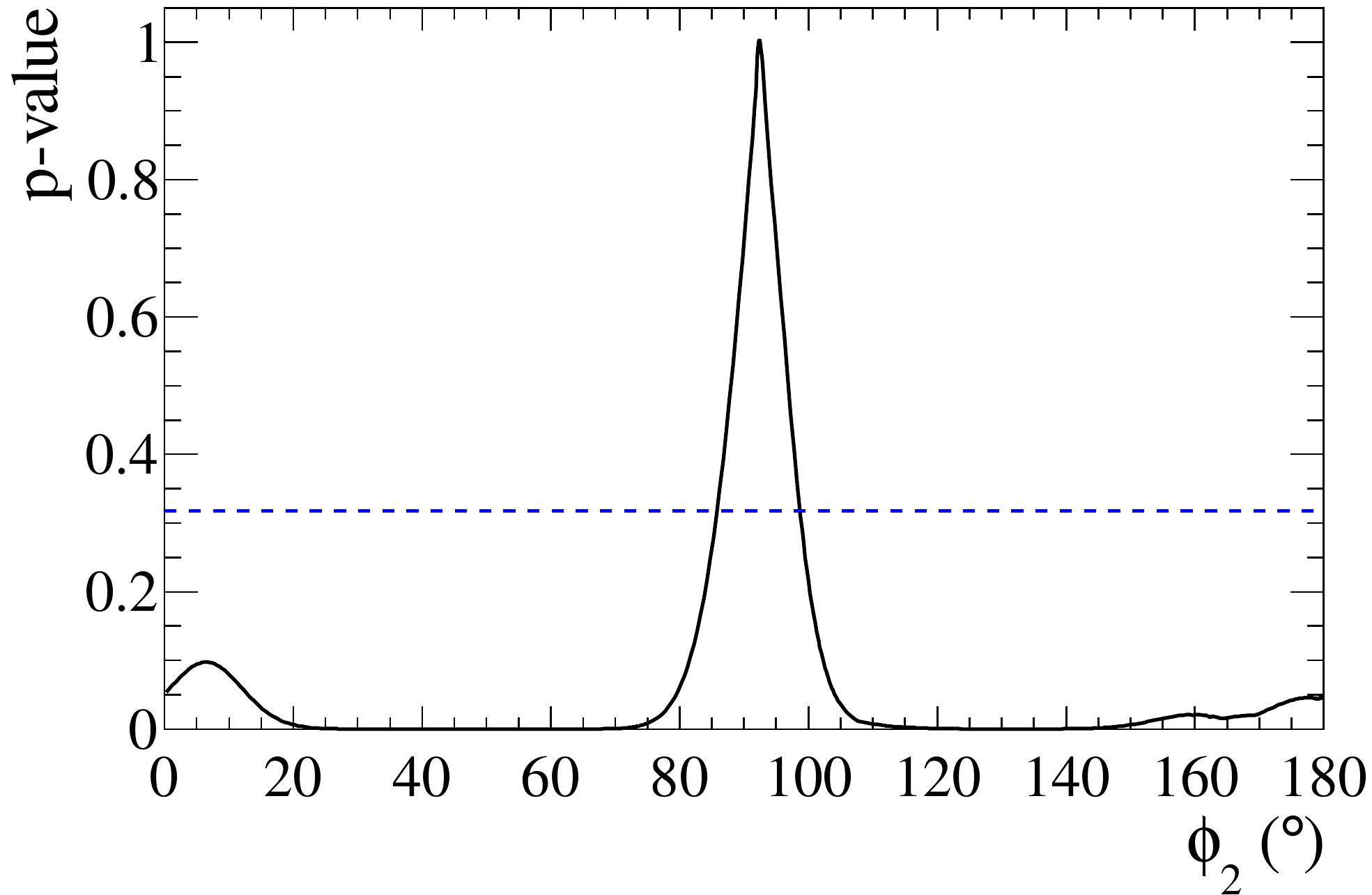}
  \includegraphics[height=120pt,width=!]{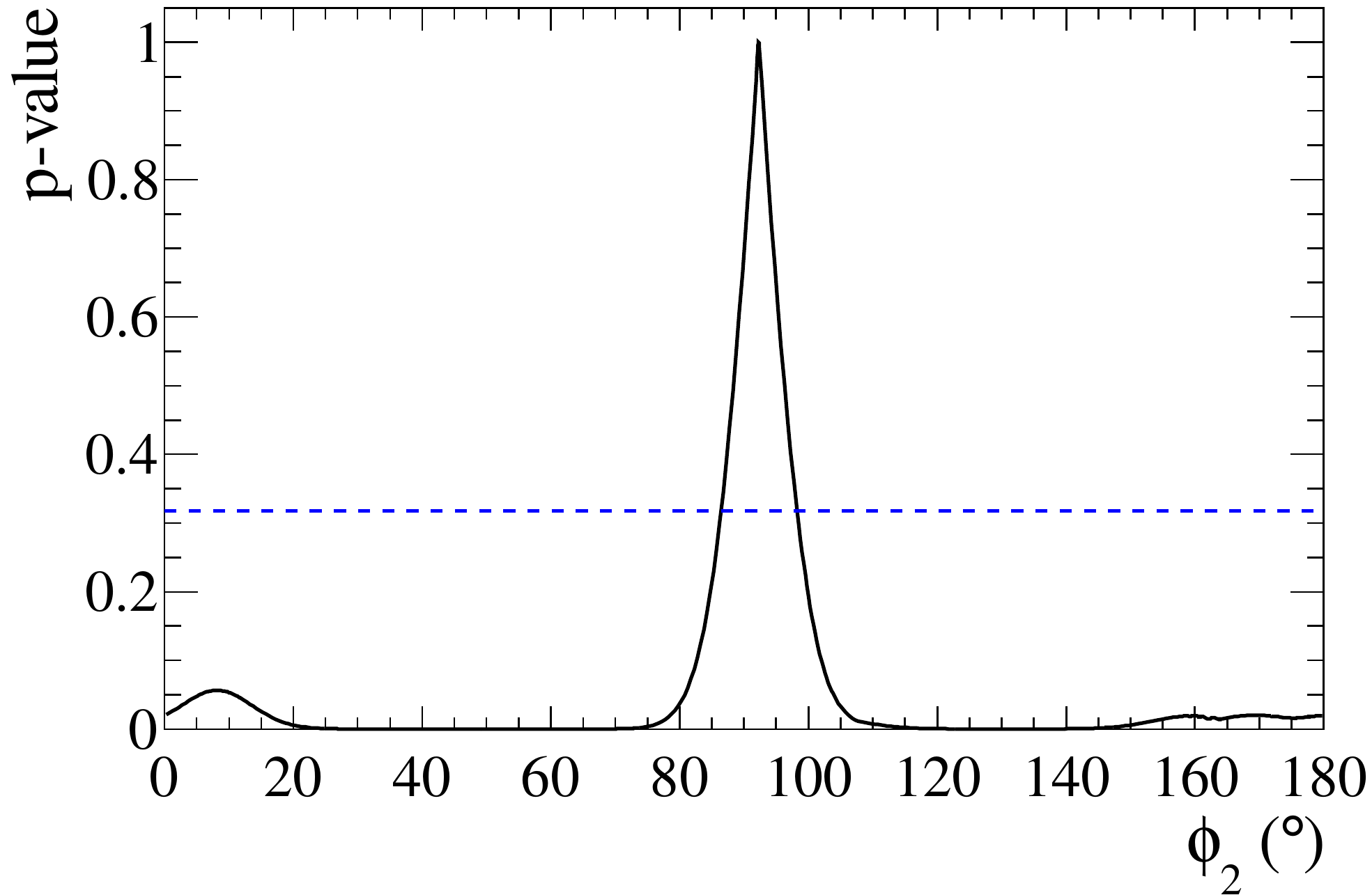}
  \put(-215,105){(c)}
  \put(-28,105){(d)}

  \includegraphics[height=120pt,width=!]{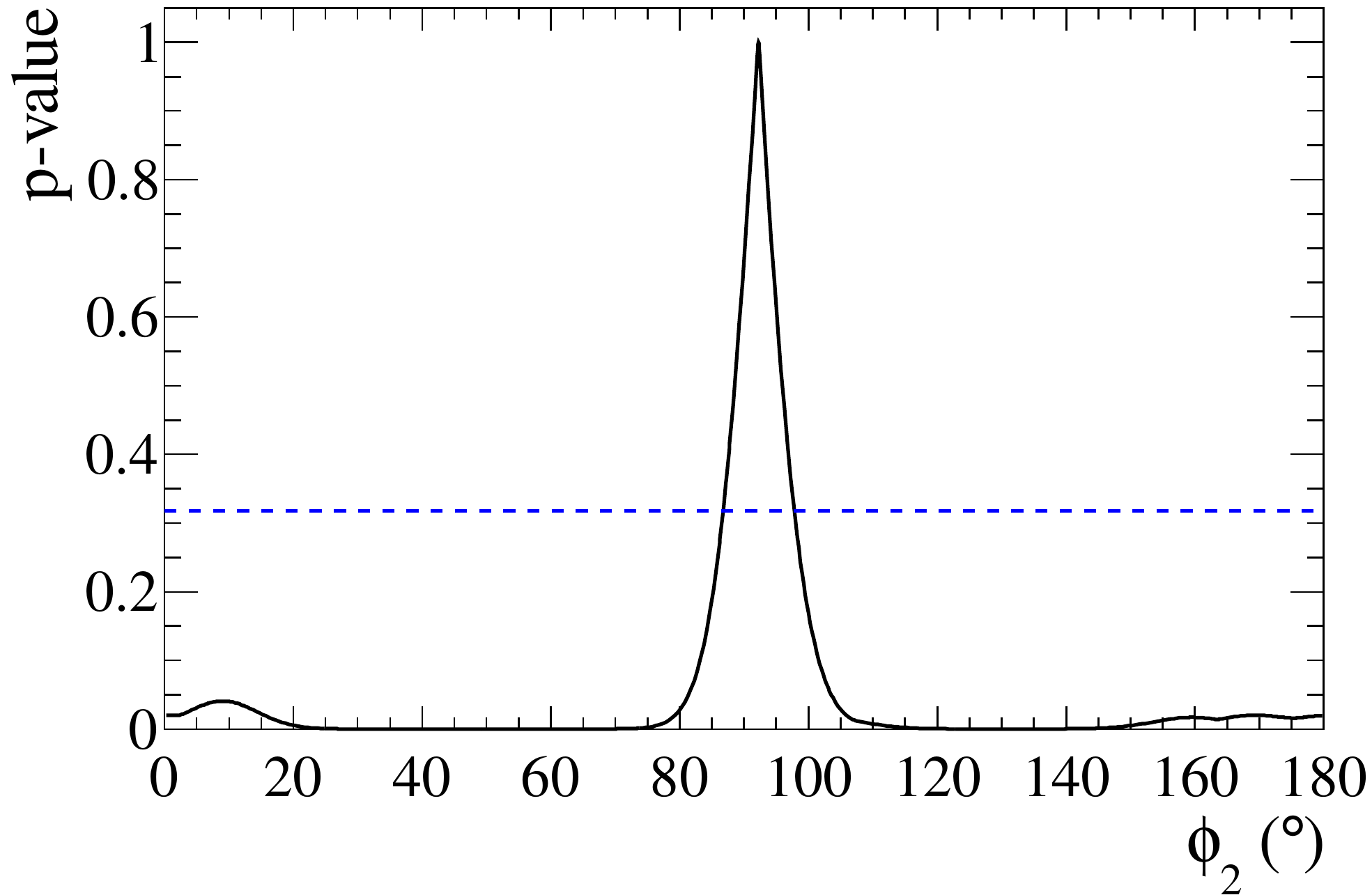}
  \includegraphics[height=120pt,width=!]{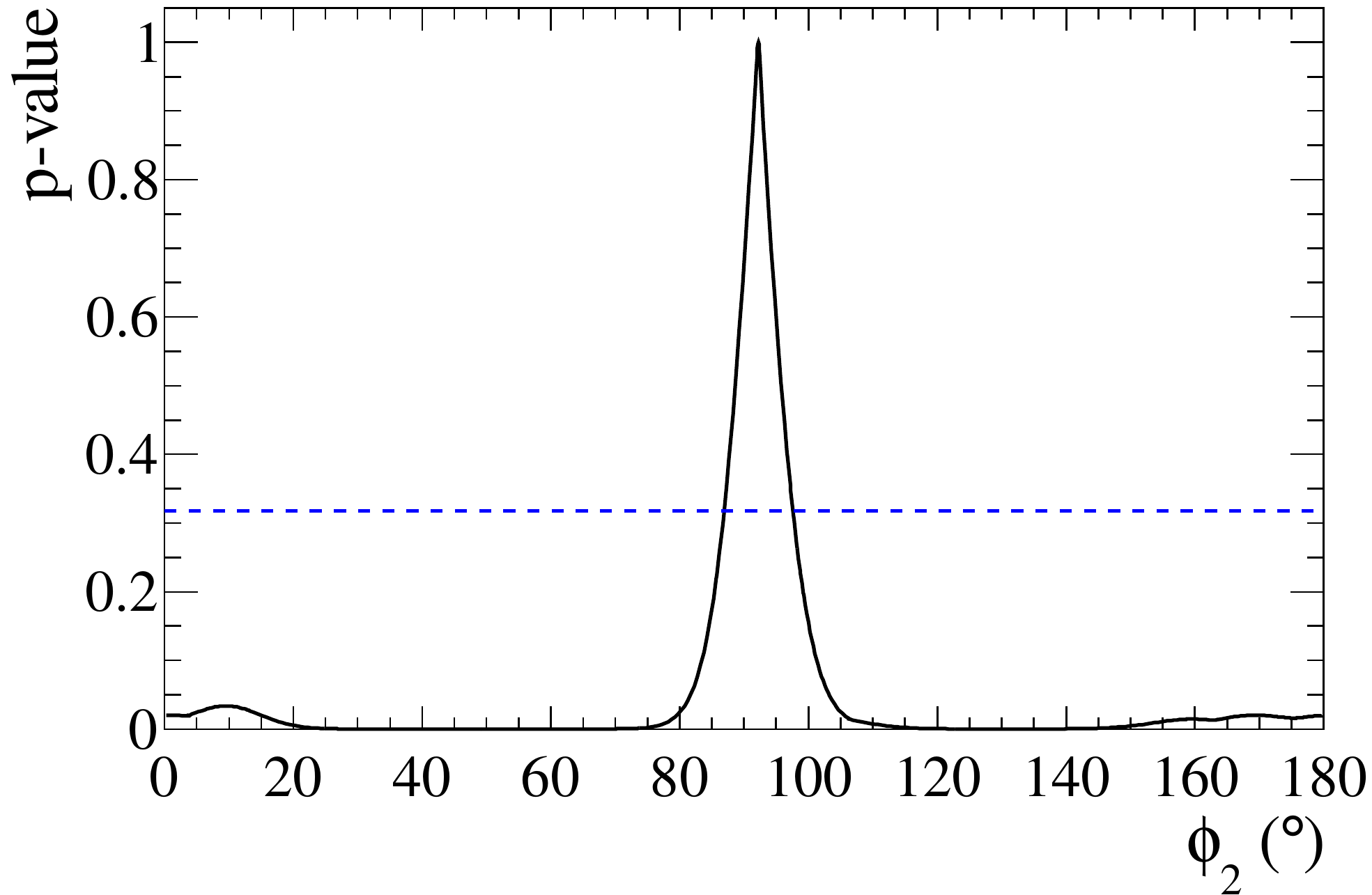}
  \put(-215,105){(e)}
  \put(-28,105){(f)}
  \caption{\label{fig:phi2_err} $p$-value scans of \phitwo\ where the horizontal dashed line shows the $1\sigma$ bound. (a) shows the standard scan involving the quasi-two-body-parameters for $\Bz \to \rhoz\rhoz$ obtained by the BaBar Collaboration. The remaining scans show the effects of increasing separation between the two solutions for $\phitwo^{00}$ from 1$\sigma$-5$\sigma$, where their uncertainties are scaled to (b) $\delta \phitwo^{00} = 50.8^\circ$ (c) $\delta \phitwo^{00} = 25.4^\circ$ (d) $\delta \phitwo^{00} = 17.0^\circ$ (e) $\delta \phitwo^{00} = 12.7^\circ$ and (f) $\delta \phitwo^{00} = 10.2^\circ$.}
\end{figure}

One issue to be aware of is that early attempts with this method may not necessarily lead to a measured $\phitwo^{00}$ that completely rules out the second solution by $5\sigma$. Particularly if this is the case, the $\chi^2$ profile of $\phitwo^{00}$ must be used instead of the usual assumption of Gaussian-distributed errors in the $\chi^2$ sum. In order to better understand the impact of increasingly preferred solutions, a test is performed on the $B \to \rho \rho$ system with the final results obtained by the BaBar Collaboration for longitudinal polarisation~\cite{phi2_rhorho1,phi2_rhorho3,phi2_rhorho5}, so chosen as they currently give the best single-experiment constraint on \phitwo~\cite{CKMFitter}. For $\phitwo^{00}$, I assume a $\chi^2$ profile that consists of the sum of $\chi^2$ terms for each solution, $\phitwo^{00} = 9^\circ$ and $\phitwo^{00} = 81^\circ$, calculated from the expression $\Scp^{00} = \sqrt{1-(\Acp^{00})^2}\sin2\phitwo^{00}$. The $\chi^2$ distribution of the first solution, which is furthest from the SM, then receives a penalty equal to the square of the statistical separation between the two solutions. In practice however, this distribution would come from a likelihood scan of $\phitwo^{00}$ convolved with the systematic uncertainty. The resulting \phitwo\ scans shown in figure~\ref{fig:phi2_err} clearly demonstrate the potential of this proposed method to resolve solutions in $B \to \rho \rho$ if the $-2\log{\cal L}$ can differentiate between $\phitwo^{00}$ solutions, as measurements of $\Scp^{00}$ would otherwise result in all scans looking like figure~\ref{fig:phi2_err}a.

\begin{figure}[tbp]
  \centering
  \includegraphics[height=120pt,width=!]{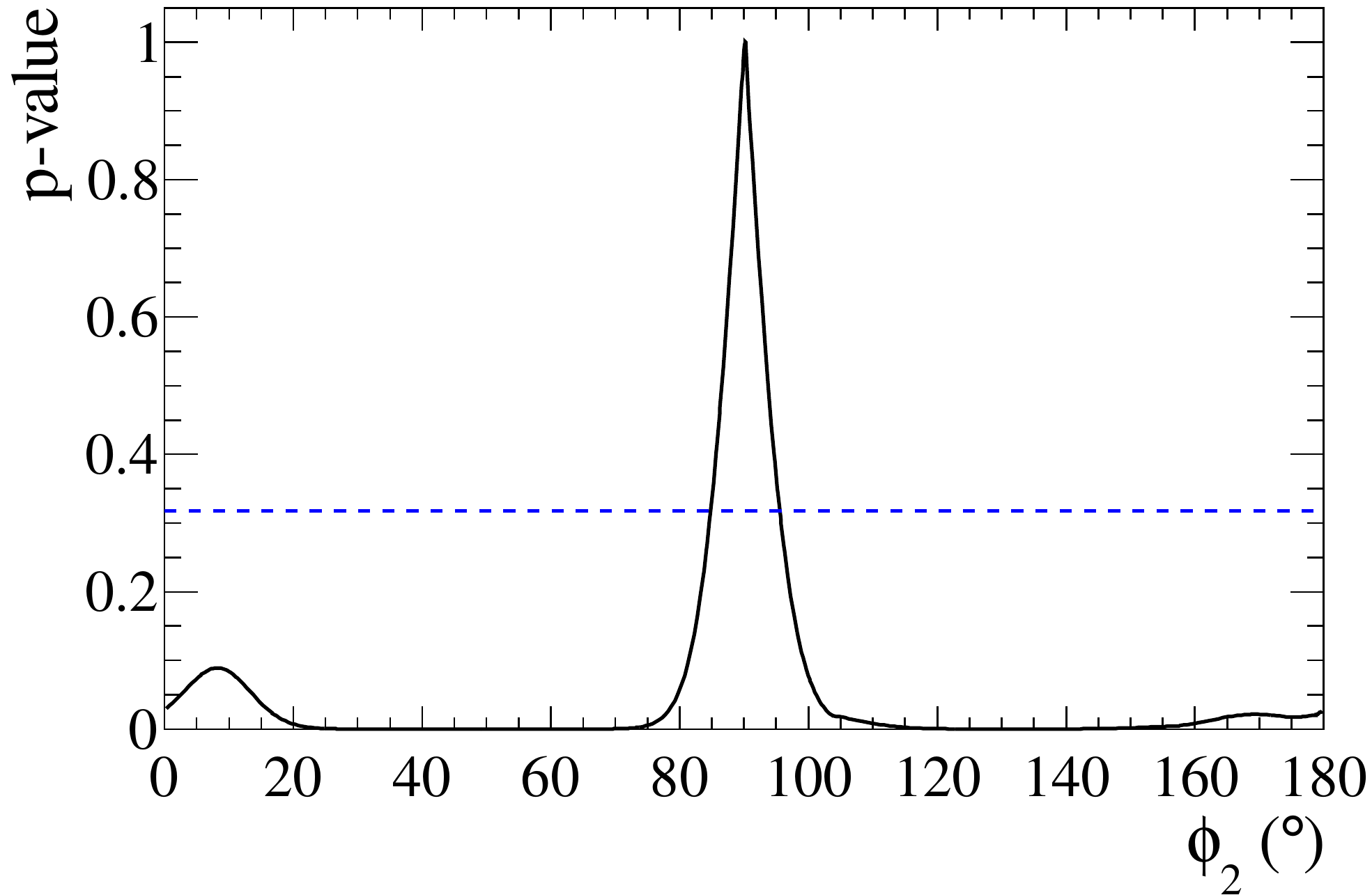}
  \includegraphics[height=120pt,width=!]{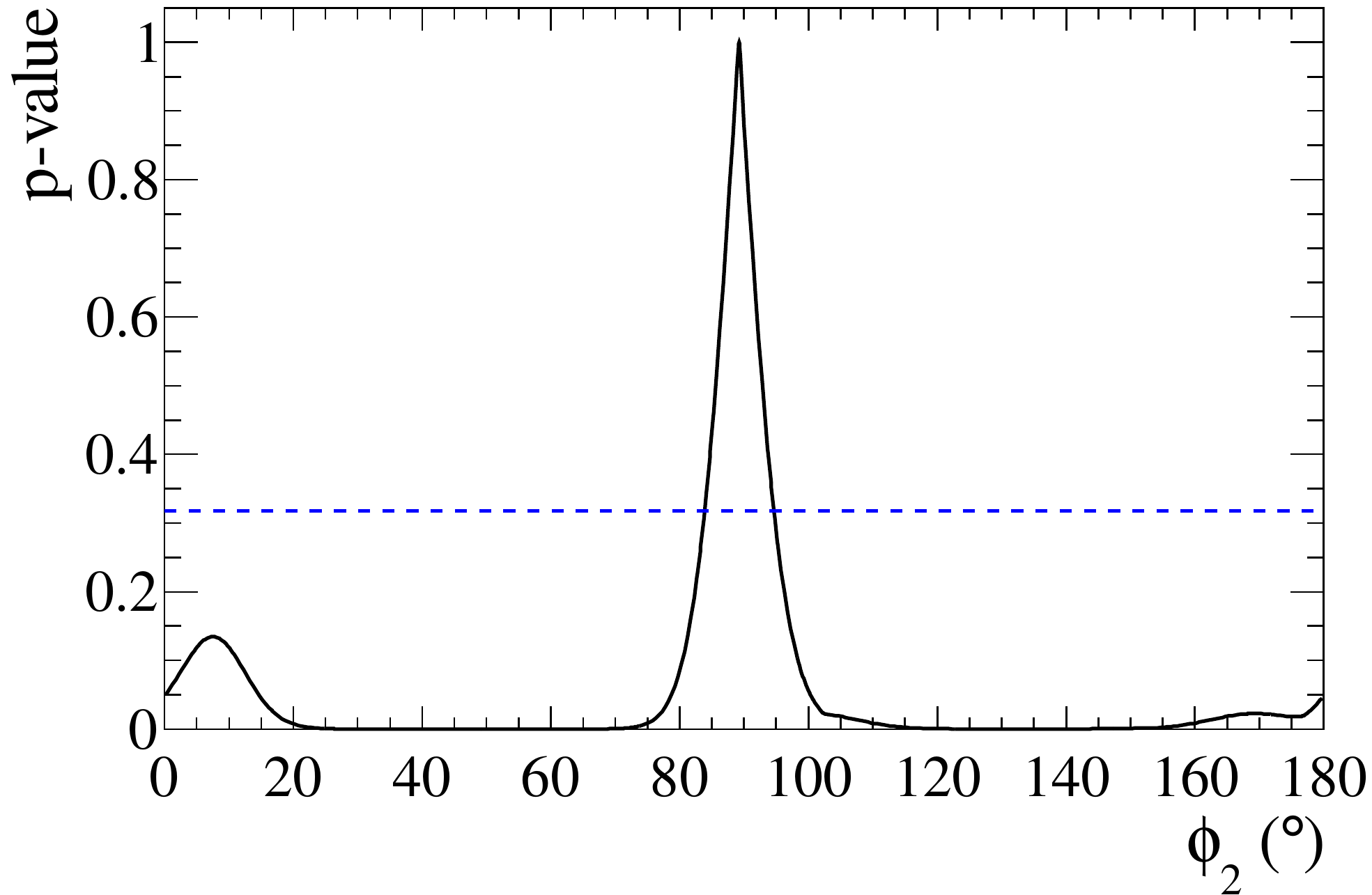}
  \put(-215,105){(a)}
  \put(-28,105){(b)}

  \includegraphics[height=120pt,width=!]{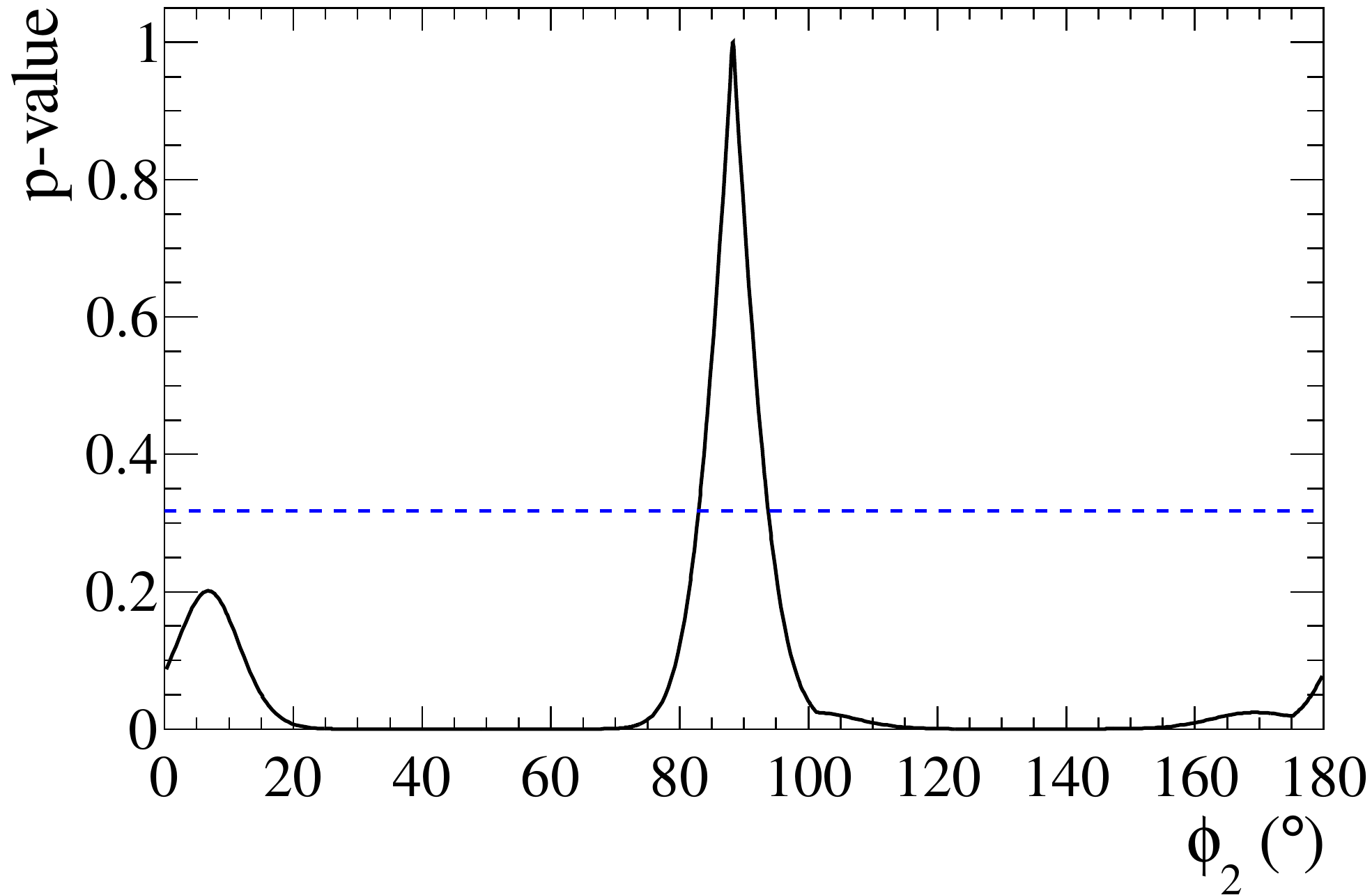}
  \includegraphics[height=120pt,width=!]{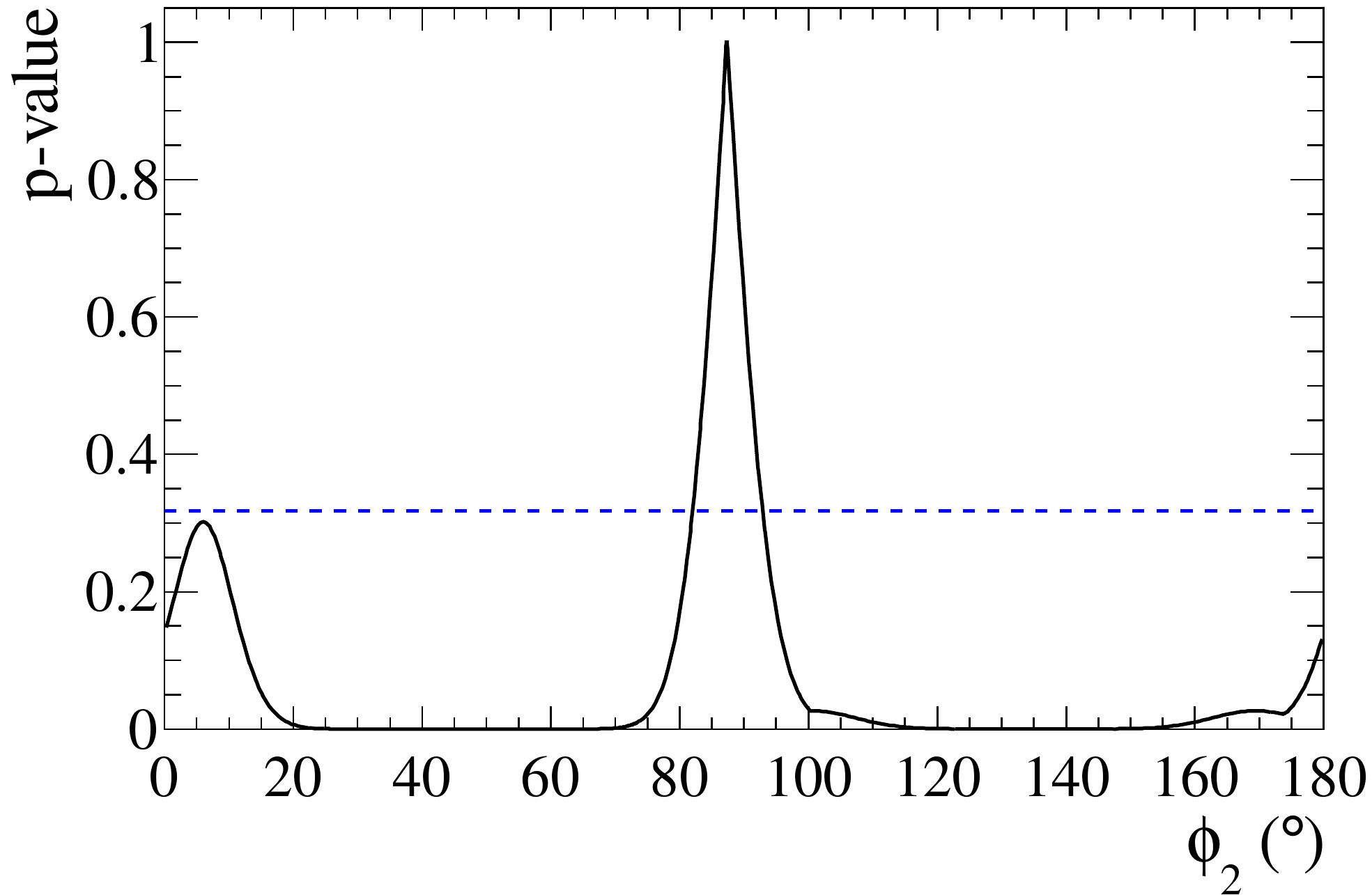}
  \put(-215,105){(c)}
  \put(-28,105){(d)}

  \includegraphics[height=120pt,width=!]{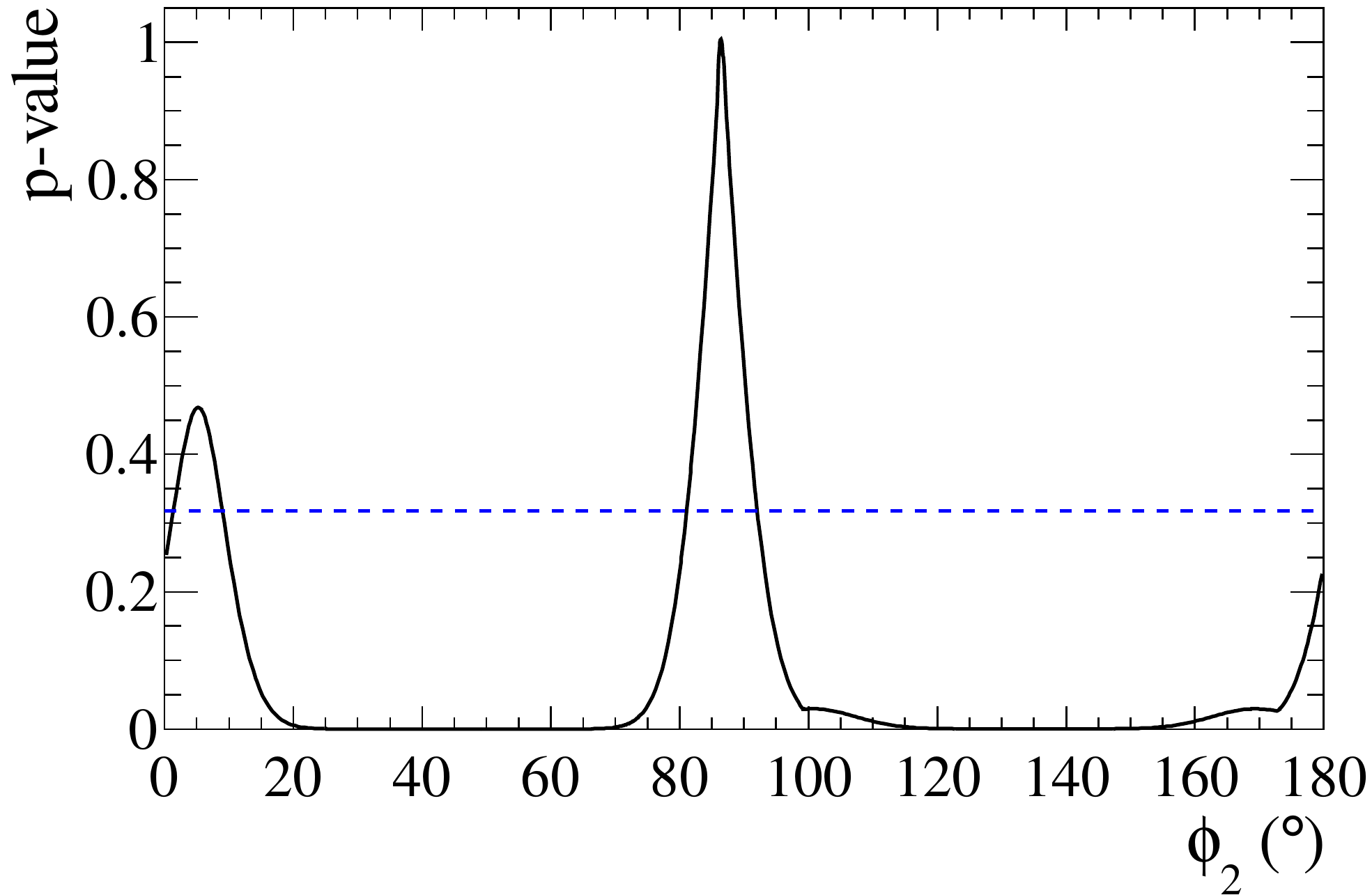}
  \includegraphics[height=120pt,width=!]{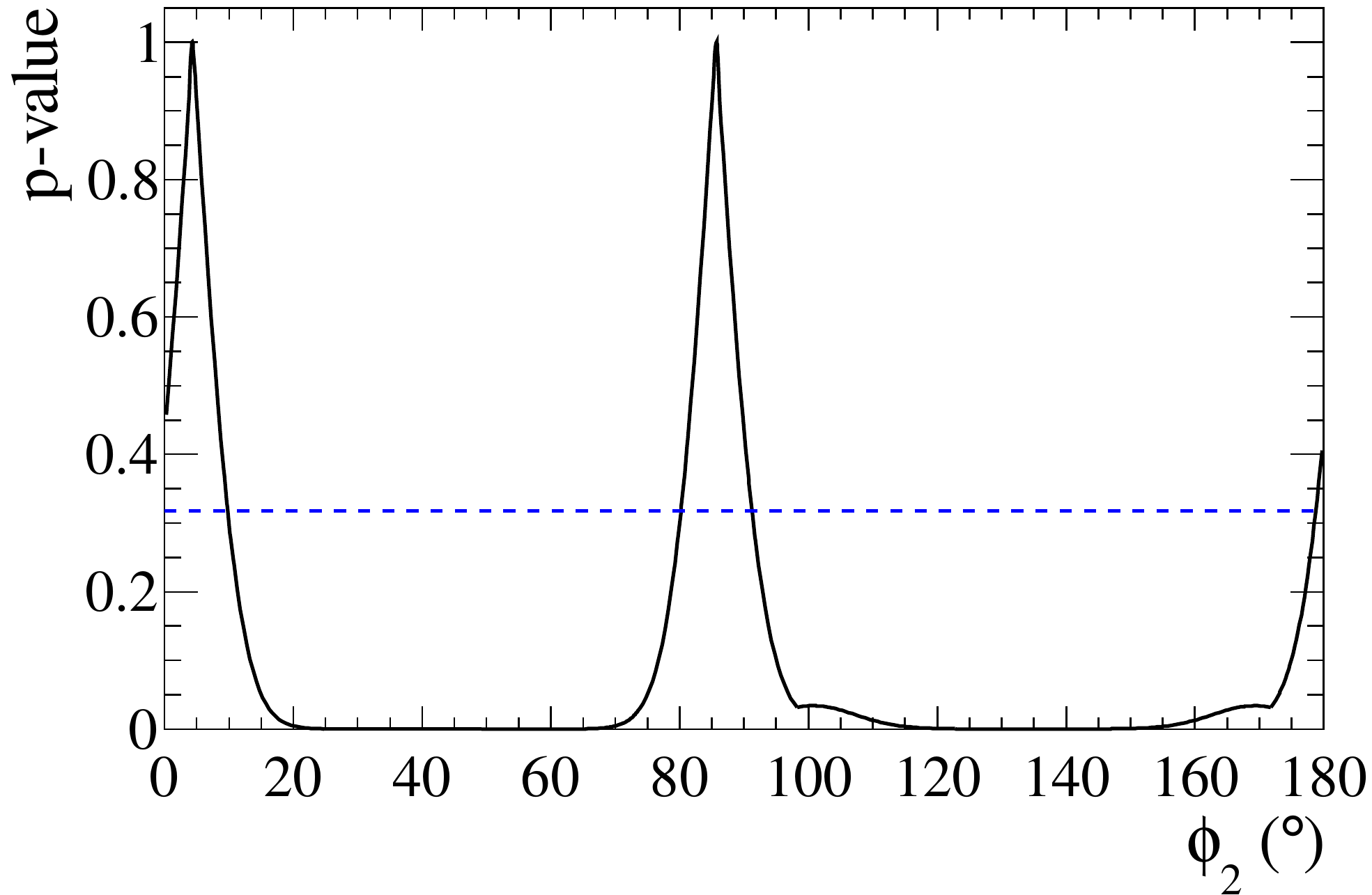}
  \put(-215,105){(e)}
  \put(-28,105){(f)}
  \caption{\label{fig:phi2_scan} $p$-value scans of \phitwo\ where the central value of $\phitwo^{00}$ is set to (a) $\phitwo^{00} = 70^\circ$ (b) $\phitwo^{00} = 65^\circ$ (c) $\phitwo^{00} = 60^\circ$ (d) $\phitwo^{00} = 55^\circ$ (e) $\phitwo^{00} = 50^\circ$ (f) $\phitwo^{00} = 45^\circ$. The horizontal dashed line indicates the $1\sigma$ bound.}
\end{figure}

It is also instructive to understand the capacity to resolve solutions in \phitwo\ as a function of the central value of $\phitwo^{00}$ itself. In this test, I assume a single solution for $\phitwo^{00}$ can be resolved to $5\sigma$ with an uncertainty of $10^\circ$, however, I allow its central value to shift. The results shown in figure~\ref{fig:phi2_scan} illustrate that when the $\phitwo^{00}$ central value lies in the vicinity between the two solutions of $\phitwo$, which are mainly driven by the precision on $\Scp^{+-}$, the resolution of its trigonometric ambiguity is not possible. If this proves to be a limiting factor in future analyses, an experimentally daunting amplitude analysis of $\Bz \to \rho^+ \rho^-$ would have to be the final resort in attempting to resolve a single solution in the $\Bz \to \rho \rho$ system.

\section{Time-dependent amplitude model}
\label{sec:model}

Now that the possibility to constrain a single solution for \phitwo\ in $B \to \rho\rho$ is established under an ideal set of circumstances, it is important to estimate the feasibility of distinguishing between the two solutions of $\phitwo^{00}$ given the current experimental knowledge of the $4\pi$ final state. A primary concern regarding this proposed method is whether the hadronic uncertainties, particularly those arising from the \aone, can be controlled at a level that will still permit a clear distinction between the two solutions. Fortunately, parameters obtained from asymmetry measurements generally tend to be more immune to the effects of the model. However, in order to demonstrate this assertion, I generate a set of pseudo-experiments varying the hadronic model within its current experimental uncertainties. I then set the yields to those expected at various milestones expected during the timelines of the Belle II and LHCb experiments and record the difference in $-2\log{\cal L}$ for each solution of $\phitwo^{00}$ obtained from a maximum likelihood fit to each pseudo-experiment. In any case, we only need sufficient precision to resolve a solution as most of the statistical power on the central value of \phitwo\ will come from $\Bz \to \rho^+ \rho^-$. Experimental effects such as detection efficiency, timing resolution and background contributions will be neglected in this study. 

\subsection{Amplitude model}
I consider a very rudimentary model with two contributions to the all-charged 4-body phase space coming only from the channels we know to exist, $\Bz \to \rhoz \rhoz$ and $\aone \pimp$. The amplitude for each intermediate state is parametrised as
\begin{equation}
  A_i(\Phi_4) = B_{L_B}(\Phi_4) \cdot [B_{L_{R_1}}(\Phi_4) T_{R_1} (\Phi_4)] \cdot [B_{L_{R_2}}(\Phi_4) T_{R_2} (\Phi_4)] \cdot S_i(\Phi_4),
\end{equation}
where $B_{L_B}$ represents the production Blatt-Weisskopf barrier factor~\cite{bwbf} depending on the orbital angular momentum between the products of the \Bz\ decays, $L_B$. Two resonances will appear in each isobar, denoted by $R_1$ and $R_2$, for which respective decay barrier factors are also assigned. The Breit-Wigner propagators are represented by $T$, while the overall spin amplitude is given by $S$. Each isobar is Bose-symmetrised so that the total amplitude is symmetric under the exchange of like-sign pions.

The Blatt-Weisskopf penetration factors account for the finite size of the decaying resonances by assuming a square-well interaction potential with radius $r$. They depend on the breakup momentum between the decay products $q$, and the orbital angular momentum between them $L$. Their explicit expressions used in this analysis are
\begin{eqnarray}
  B_0(q) &=& 1, \nonumber \\
  B_1(q) &=& \frac{1}{\sqrt{1+(qr)^2}}, \nonumber\\
  B_2(q) &=& \frac{1}{\sqrt{9+3(qr)^2+(qr)^4}}.
\end{eqnarray}

In general, resonance lineshapes are described by Breit-Wigner propagators as a function of the energy-squared $s$,
\begin{equation}
  T(s) = \frac{1}{M^2(s) - s - im_0\Gamma(s)},
\end{equation}
where $M^2(s)$ is the energy-dependent mass and $\Gamma(s)$ is the total width which is normalised such that it represents the nominal width $\Gamma_0$ at the pole mass $m_0$.

\begin{figure}[tbp]
  \centering
  \includegraphics[height=120pt,width=!]{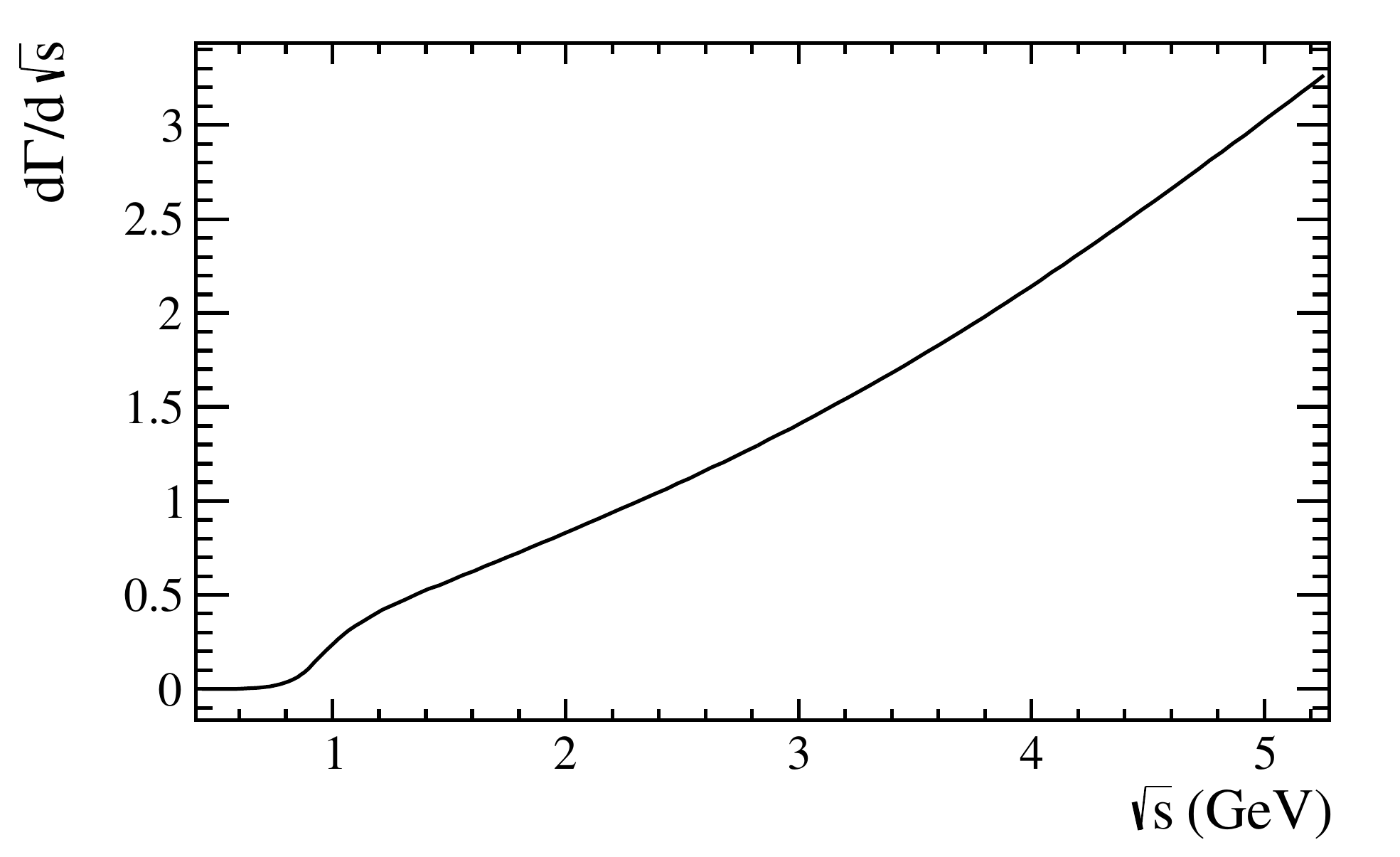}
  \caption{\label{fig:a1} Energy-dependent width of the \aone.}
\end{figure}

For the \rhoz\ resonance, the Gounaris-Sakurai parametrisation is used to provide an analytic expression for $M^2(s)$ and $\Gamma(s)$~\cite{gs}. I ignore dispersive effects in the \aone, setting $M^2(s)$ to the pole-mass squared. The energy-dependent width of the \aone\ is calculated from the integral of its Dalitz Plot as a function of $s$,
\begin{equation}
  \Gamma_{\aone}(s) = \frac{1}{2\sqrt{s}} \int |A_{\aone \to (\rho\pi)^\pm_S} (s)|^2 d\Phi_3,
\end{equation}
where exclusive decay to $(\rho\pi)^\pm$ in an $S$-wave configuration and isospin symmetry i.e., $\Gamma_{\aone \to \rhoz \pipm}(s) = \Gamma_{\aone \to \rho^\pm \piz}(s)$ is also assumed. The form of the \aone\ energy-dependent width can be seen in figure~\ref{fig:a1}.

Spin amplitudes are constructed with the covariant tensor formalism based on the Rarita-Schwinger conditions~\cite{spin}. The spin $S$, of some state with 4-momentum $p$ and spin projection $s_z$, is represented by a rank-$S$ polarisation tensor that is symmetric, traceless and orthogonal to $p$. These conditions reduce the number of independent elements to $2S+1$ in accordance with the number of degrees of freedom available to a spin-$S$ state. The sum over these polarisation indices of the inner product of polarisation tensors form the fundamental basis on which all spin amplitudes are built. Called the spin projection operator $P$, it has the capacity to project an arbitrary tensor onto the subspace spanned by the spin projections of the spin-$S$ state.

Another particularly useful object is the relative orbital momentum spin tensor $L$, which for some process $R \to AB$, is the relative momenta of the decay products $q_R \equiv p_A - p_B$ projected to align with the spin of $R$,
\begin{equation}
  L_{\mu_1 \mu_2 ... \mu_L}(p_R, q_R) = P_{\mu_1 \mu_2 ... \mu_L \nu_1 \nu_2 ... \nu_L} (p_R) q_R^{\nu_1} q_R^{\nu_2} ... q_R^{\nu_L}.
\end{equation}
Finally, to ensure that the spin amplitude behaves correctly under parity transformation, it is sometimes necessary to include the Levi-Cevita totally antisymmetric tensor $\epsilon_{abcs}p_R^d$. Each stage of a decay is represented by a Lorentz scalar obtained by contracting an orbital tensor between the decay products with a spin wavefunction of equal rank representing the final state.

Four topologies are necessary for this analysis. They include three for $\Bz \to \rhoz \rhoz$ as $S$, $P$ and $D$ waves are permitted between the two \rhoz\ resonances,
\begin{eqnarray}
  &&S\text{-wave}: \hspace{10pt} S \propto L_a(p_{\rhoz_1}, q_{\rhoz_1})L^a(p_{\rhoz_2}, q_{\rhoz_2}), \nonumber\\
  &&P\text{-wave}: \hspace{10pt} S \propto \epsilon_{abcd} L^d(p_{\Bz}, q_{\Bz}) L^c(p_{\rhoz_1}, q_{\rhoz_1}) L^b(p_{\rhoz_2}, q_{\rhoz_2}) p^a_{\Bz}, \nonumber\\
  &&D\text{-wave}: \hspace{10pt} S \propto L_{ab}(p_{\Bz}, q_{\Bz}) L^b(p_{\rhoz_1}, q_{\rhoz_1}) L^a(p_{\rhoz_2}, q_{\rhoz_2}).
\end{eqnarray}
For $\Bz \to \aone \pimp$, I consider only a single topology arising from the $S$-wave between the products of the $\aone \to \rho^0 \pipm$ decay,
\begin{equation}
  S \propto L_a(p_{\Bz}, q_{\Bz}) P^{ab}(p_{\aone}) L_b(p_{\rhoz}, q_{\rhoz}),
\end{equation}
as the final BaBar and Belle analyses measured signal yields with a dipion selection criteria optimised in order to isolate the $\rho^0$~\cite{phi2_a1pi1,phi2_a1pi2}. However, future analyses will also have to consider the overlapping $\aone \to \sigma \pipm$ channel which contributes $\sim 25\%$ of the $\aone \to \rho^0 \pipm$ rate~\cite{CLEO_4pi}, leading to larger $4\pi$ yields than those estimated here.

\subsection{Pseudo-experiment generation method}

In order to get some early idea on the impact of hadronic and $CP$ violating uncertainties on the $\phitwo^{00}$ measurement, I adopt a procedure of varying these within current experimental uncertainties for each pseudo-experiment. The average uncertainty on $\phitwo^{00}$ can then be interpreted as the quadratic sum of the expected statistical error with the sources of model uncertainty considered in this analysis. The Monte Carlo (MC) is based on eq.~\ref{eq:tdrate} and requires a 2-stage process for generation. The first stage sets the model parameters of the $\Bz \to \rhoz \rhoz$ and $\Bz \to \aone \pimp$ isobars individually. A second stage is required to reverse engineer the complex couplings between them.

Table~\ref{tab:stage1} records the generated parameters needed for stage 1. For $\Bz \to \rhoz \rhoz$, the $S$ and $D$-waves between the two \rhoz\ mesons are practically indistinguishable in the phase space in stark contrast to $D^0$ decays, where the small size of phase space destroys some of the Bose-symmetry responsible for this degeneracy appearing in \Bz\ decays. This means that an amplitude analysis is only sensitive to the component of the $D$-wave that is orthogonal to the $S$-wave, so the real part of the $D$-wave is fixed to zero as it will anyway be absorbed by the $S$-wave amplitude which is fixed along the real axis. The $P$-wave is completely distinguishable and so a random phase is assigned in this stage. The $CP$ violation parameters of $\Bz \to \rhoz \rhoz$ are shared between the 3 polarisations. I assume direct $CP$ violation and a $\phitwo^{00}$ solution closest to the SM inferred by the BaBar measurement~\cite{phi2_rhorho5}.

\begin{table}[tbp]
  \centering
  \begin{tabular}{|c|c|c|}
    \hline 
    Parameter & Range & Reference\\ \hline
    $\tau_{\Bz}$ & $1.520 \pm 0.004$ ps  & \cite{PDG}\\
    $\Delta m_d$ & $0.5064 \pm 0.0019$ ps${}^{-1}$ & \cite{PDG}\\

    $r$ & $[2, 6]$ $c/$GeV & ---\\
    $m_0(\rhoz)$ & $0.7690 \pm 0.0009$ GeV$/c^2$ & \cite{PDG}\\
    $\Gamma_0(\rhoz)$ & $0.1509 \pm 0.0017$ GeV & \cite{PDG}\\
    $\arg(a_{\Bz \to (\rhoz \rhoz)_P})$ & $[-180, +180]^\circ$ & ---\\
    $|\lambda^{00}|$ & $0.8$ & \cite{phi2_rhorho5}\\
    $\phitwo^{00}$ & $81^\circ$ & \cite{phi2_rhorho5}\\

    $m_0(\aone)$ & $1.225 \pm 0.022$ GeV$/c^2$ & \cite{CLEO_4pi}\\
    $\Gamma_0(\aone)$ & $0.430 \pm 0.039$ GeV & \cite{CLEO_4pi}\\
    $\arg(a_{\Bz \to a_1^+ \pim})$ & $[-180, +180]^\circ$ & ---\\
    $\phitwo(\Bz \to a_1^+ \pim)$ & $(97.2 \pm 9.3)^\circ$ & \cite{a1pi_th2,phi2_a1pi2}\\
    $\phitwo(\Bz \to a_1^- \pip)$ & $(107.0 \pm 16.9)^\circ$ & \cite{a1pi_th2,phi2_a1pi2}\\
    $\Acp(\Bz \to \aone \pimp)$ & $-0.06 \pm 0.09$ & \cite{phi2_a1pi2}\\
    $\Ccp(\Bz \to \aone \pimp)$ & $-0.01 \pm 0.14$ & \cite{phi2_a1pi2}\\
    $\Delta {\cal C}(\Bz \to \aone \pimp)$ & $+0.54 \pm 0.13$ & \cite{phi2_a1pi2}\\
    $\Delta {\cal S}(\Bz \to \aone \pimp)$ & $-0.09 \pm 0.15$ & \cite{phi2_a1pi2}\\
    \hline 
  \end{tabular}
  \caption{Stage 1 parameters. Uncertainties indicate the parameter was Gaussian distributed, square brackets indicate uniform generation within the range while a single value is a constant of generation.}
  \label{tab:stage1}
\end{table}

For $\Bz \to \aone \pimp$, the relative phase to the fixed $\Bz \to \rhoz \rhoz$ $S$-wave is unknown, so it is also assigned randomly in the first stage. Certain amplitude-level parameters can be derived from the measured quasi-two-body parameters of $\Bz \to \aone \pimp$ listed in the final four entries of table~\ref{tab:stage1}, for which the Belle result is selected~\cite{phi2_a1pi2}. They include the relative magnitude of $\Bz \to a_1^- \pip$ compared to the dominant $\Bz \to a_1^+ \pim$, which is given by
\begin{equation}
  |a_{\Bz \to a_1^- \pip}|/|a_{\Bz \to a_1^+ \pim}| = \sqrt{\frac{1-\Delta{\cal C}}{1+\Delta{\cal C}}},
\end{equation}
and their relative phase difference given by
\begin{equation}
  \arg(a_{\Bz \to a_1^- \pip}/a_{\Bz \to a_1^+ \pim}) = \arcsin (\Delta{\cal S}).
\end{equation}
A coin-flip determines which solution for the relative phase is chosen in each pseudo-experiment.

The magnitude of the $\Bz \to \aone \pimp$ $CP$ violation parameters can also be determined from
\begin{equation}
  |\lambda(\Bz \to \aone \pimp)| = \sqrt{\frac{1+A_{\mp}}{1-A_{\mp}}},
\end{equation}
where
\begin{equation}
  A_+ = \frac{\Acp - \Ccp - \Acp\Delta{\cal C}}{1 - \Delta{\cal C} - \Acp\Ccp}, \hspace{10pt} A_- = -\frac{\Acp + \Ccp + \Acp\Delta{\cal C}}{1 + \Delta{\cal C} + \Acp\Ccp},
\end{equation}
which represents the direct $CP$ violation in $\Bz \to a_1^- \pip$ and $\Bz \to a_1^+ \pim$, respectively. For $\Bz \to \aone \pimp$, \Acp\ refers to the time and flavour-integrated $CP$ violation, while \Ccp\ is the flavour-dependent direct $CP$ violation.

The remaining parameters are the effective \phitwo\ phases of the two $\Bz \to \aone \pimp$ channels. These cannot be inferred from the quasi-two-body results obtained from existing time-dependent analyses. However, due to the remarkable agreement between the Belle result and the predictions from QCD factorisation~\cite{a1pi_th2}, I select central values based on their theoretical prediction. The uncertainties in contrast, are taken from the Belle measurement which is sensitive to their algebraic average, inflated to reflect the size of $\Delta{\cal C}$, which is a measure of the relative difference in rates between $\Bz \to a_1^+ \pim$ and $\Bz \to a_1^- \pip$.

The remaining amplitude-level parameters are obtained in the second stage of parameter generation. As two strong phases were uniformly assigned in stage 1, their associated magnitudes have to be calculated. The imaginary part of the $D$-wave $\Bz \to \rhoz \rhoz$ also needs to be determined as its real part is absorbed into the $S$-wave. These can be reverse-engineered from the parameters listed in table~\ref{tab:stage2}.

\begin{table}[tbp]
  \centering
  \begin{tabular}{|c|c|c|}
    \hline 
    Parameter & Range & Reference\\ \hline
    ${\cal B}(\Bz \to \rhoz \rhoz)$ & $(0.95 \pm 0.16)\times10^{-6}$ & \cite{HFAG}\\
    $f_L(\Bz \to \rhoz \rhoz)$ & $0.745 \pm 0.067$ & \cite{phi2_rhorho7}\\
    $f^\prime_\parallel(\Bz \to \rhoz \rhoz)$ & $0.5 \pm 0.1$ & \cite{phi2_rhorho7}\\
    $f_S(\Bz \to \rhoz \rhoz)$ & $[0.5, 1.0][1-f_P(\Bz \to \rhoz \rhoz)]$ & ---\\
    ${\cal B}(\Bz \to \aone \pimp){\cal B}(\aone \to \pipm \pip \pim)$ & $(11.1 \pm 1.7)\times10^{-6}$ & \cite{phi2_a1pi2}\\
    \hline 
  \end{tabular}
  \caption{Stage 2 parameters. Uncertainties indicate the parameter was Gaussian distributed while square brackets indicate uniform generation within the range.}
  \label{tab:stage2}
\end{table}

In order to generate the $\Bz \to \rhoz \rhoz$ branching fractions for each polarisation, the combined branching fraction is first generated. I then take the LHCb measurement of the longitudinal and relative parallel polarisation~\cite{phi2_rhorho7} which immediately sets the $P$-wave fraction through
\begin{equation}
  f_P(\Bz \to \rhoz \rhoz) = [1-f_L(\Bz \to \rhoz \rhoz)] f^\prime_\parallel(\Bz \to \rhoz \rhoz),
\end{equation}
according to the definition set by LHCb. For the remaining branching fractions of the $S$ and $D$-waves, the split between them is uniformly generated where the $S$-wave is assumed to dominate.

To generate the branching fractions of $\Bz \to a_1^+ \pim$ and $\Bz \to a_1^- \pip$, the combined branching fraction is generated from the Belle result~\cite{phi2_a1pi2}, where the relative amounts can be determined from parameters already generated in stage 1,
\begin{equation}
  {\cal B}(\Bz \to a_1^+ \pim){\cal B}(\aone \to \pipm \pip \pim) =  \frac{1 + \Delta{\cal C} + \Acp\Ccp}{2} {\cal B}(\Bz \to \aone \pimp){\cal B}(\aone \to \pipm \pip \pim).
\end{equation}

The remaining 3 amplitude-level parameters of the model are then determined from a $\chi^2$ fit relating the generated branching fractions for each isobar scaled to unity, to the fit fractions of each isobar calculated for the generated model in the 4-body phase space,
\begin{equation}
  {\cal F}_i = \frac{\int (|A_i|^2 + |\bar A_i|^2) d\Phi_4}{\int \sum_i(|A_i|^2 + |\bar A_i|^2) d\Phi_4}.
\end{equation}

\subsection{Expected yields}
Estimates of the size of future event samples can be obtained by extrapolating from existing results. Belle has presented results on $\Bz \to \rhoz \rhoz$ and ${\cal B}(\Bz \to \aone \pimp)$ based on their final data sample of 772 million $B \bar B$ pairs collected at the $\Upsilon(4S)$ resonance~\cite{phi2_rhorho6, phi2_a1pi2}, while LHCb has performed a study of $\Bz \to \rhoz \rhoz$, exploiting the full 3 fb${}^{-1}$ Run 1 data set~\cite{phi2_rhorho7}. I generate pseudo-experiments based on individual amplitude-models with the \verb|GENBOD| phase space function~\cite{phsp} and \verb|qft++| to provide the spin densities~\cite{qft}. Four tests are performed with crude estimates of the combined $\Bz \to \rhoz \rhoz$ and ${\cal B}(\Bz \to \aone \pimp)$ effective yields for 10 ab$^{-1}$ and the full 50 ab$^{-1}$ of data expected with Belle II~\cite{belle2}, as well as the amount of data expected to be recorded by LHCb at the end of Run 2~\cite{run2} and Run 3~\cite{run3}. These are recorded in table~\ref{tab:yields}.

Belle found a combined total yield of around 1600 events. Neglecting efficiency differences and assuming improvements in selection arising from the widened phase space analysis region and better detector, a yield of around 90000 events should be realistic at Belle II. With a conservative estimate of 30\% for the effective tagging efficiency, approximately 30000 effective signal events would remain.

In their analysis of $\Bz \to \rhoz \rhoz$, LHCb suppressed $\Bz \to \aone \pimp$ contributions to negligible levels in exchange for 50\% loss of signal, by requiring every 3-pion combination to have an invariant mass in excess of 2.1~GeV/$c^2$~\cite{phi2_rhorho7}. This overlap region is critical to the method of attempting to extract a single solution for $\phitwo^{00}$, thus I project a yield of 1000 $\Bz \to \rhoz \rhoz$ events could be achieved by opening up the phase space to include $\Bz \to \aone \pimp$ in Run 1. Again assuming similar detection efficiencies for $\Bz \to \aone \pimp$, the combined yield would be near 12000 events. If Run 2 can achieve twice this yield, an effective tagging efficiency of 4\% would leave an effective yield of around 1500 events. Run 3 is roughly estimated to procure an order of magnitude more data over that expected by naively scaling expected integrated luminosities, due to the planned removal of the hardware (L0) trigger and migration to a fully software-based system~\cite{trigger}.

\begin{table}[tbp]
  \centering
  \begin{tabular}{|c|c|c|c|}
    \hline 
    Experiment & Milestone & Year & Effective Yield\\ \hline
    LHCb & 8 fb$^{-1}$ (Run 2) & 2018 & 1500\\
    Belle II & 10 ab$^{-1}$ & 2021 & 6000\\
    LHCb & 23 fb$^{-1}$ (Run 3) & 2023 & 15000\\
    Belle II & 50 ab$^{-1}$ & 2024 & 30000\\
    \hline 
  \end{tabular}
  \caption{Expected combined effective tagged yields of $\Bz \to \rhoz \rhoz$ and $\Bz \to \aone \pimp$.}
  \label{tab:yields}
\end{table}

\section{Results}
\label{sec:results}

\begin{figure}[tbp]
  \centering
  \includegraphics[height=120pt,width=!]{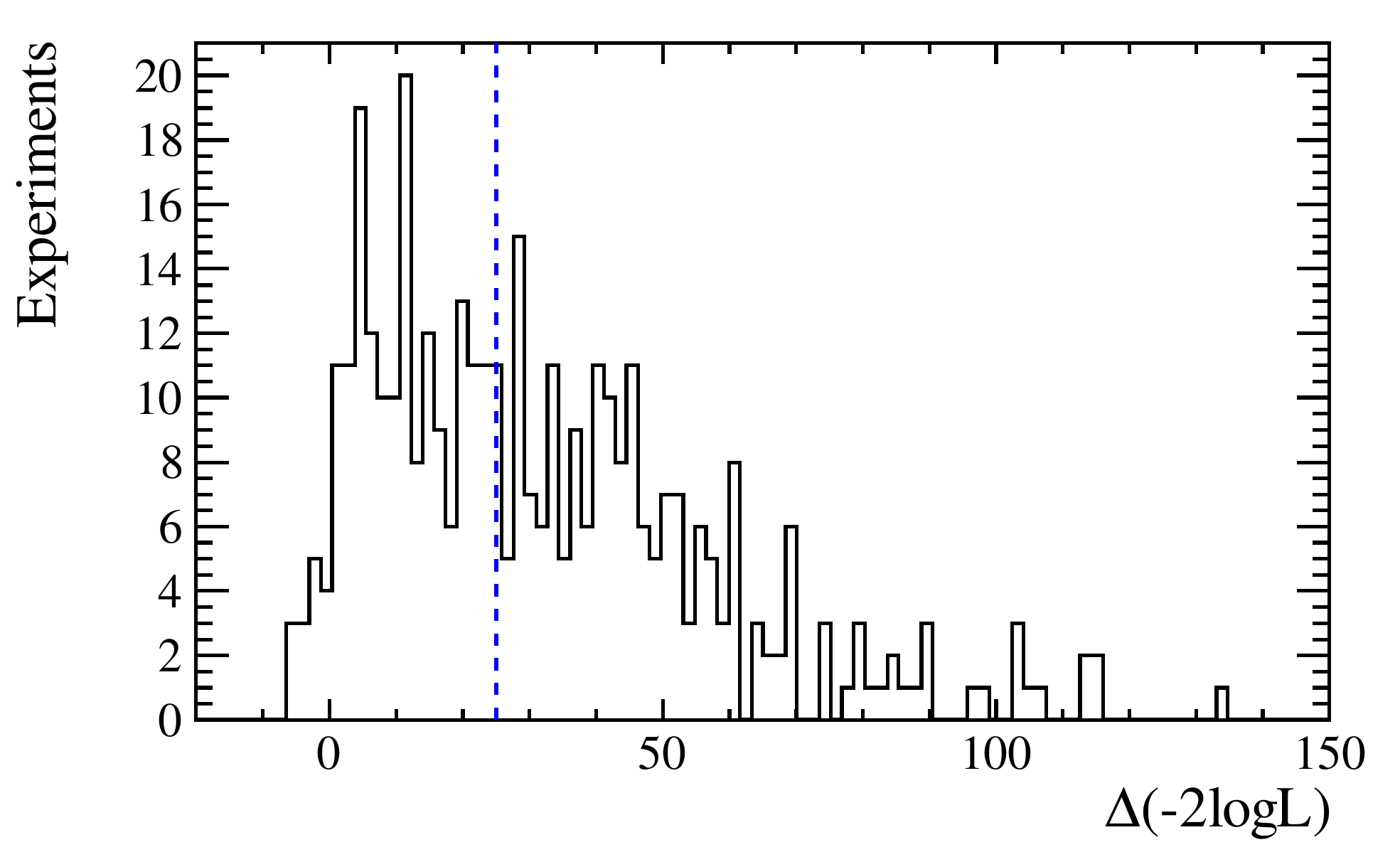}
  \includegraphics[height=120pt,width=!]{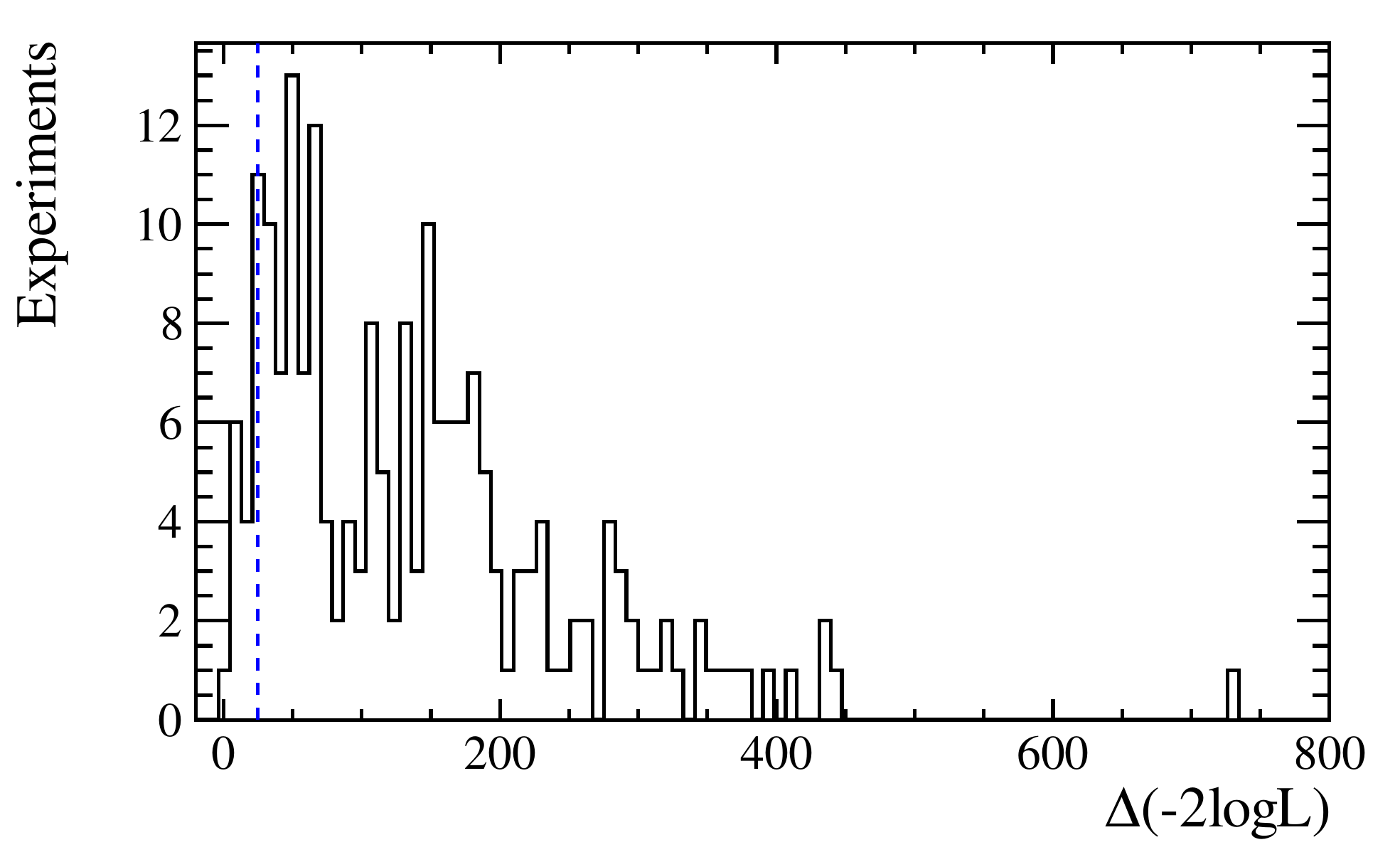}
  \put(-230,100){(a)}
  \put(-32,100){(b)}

  \includegraphics[height=120pt,width=!]{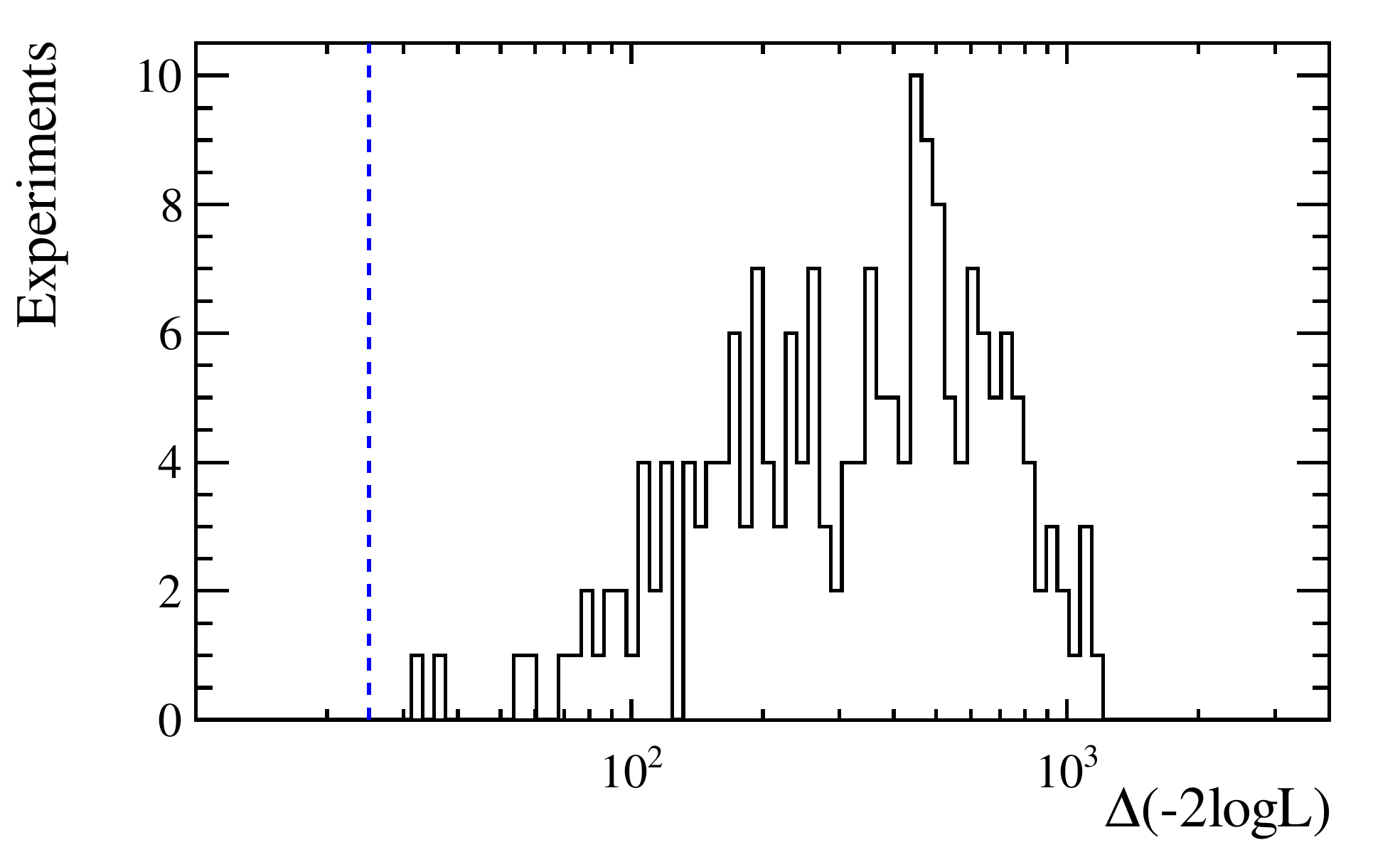}
  \includegraphics[height=120pt,width=!]{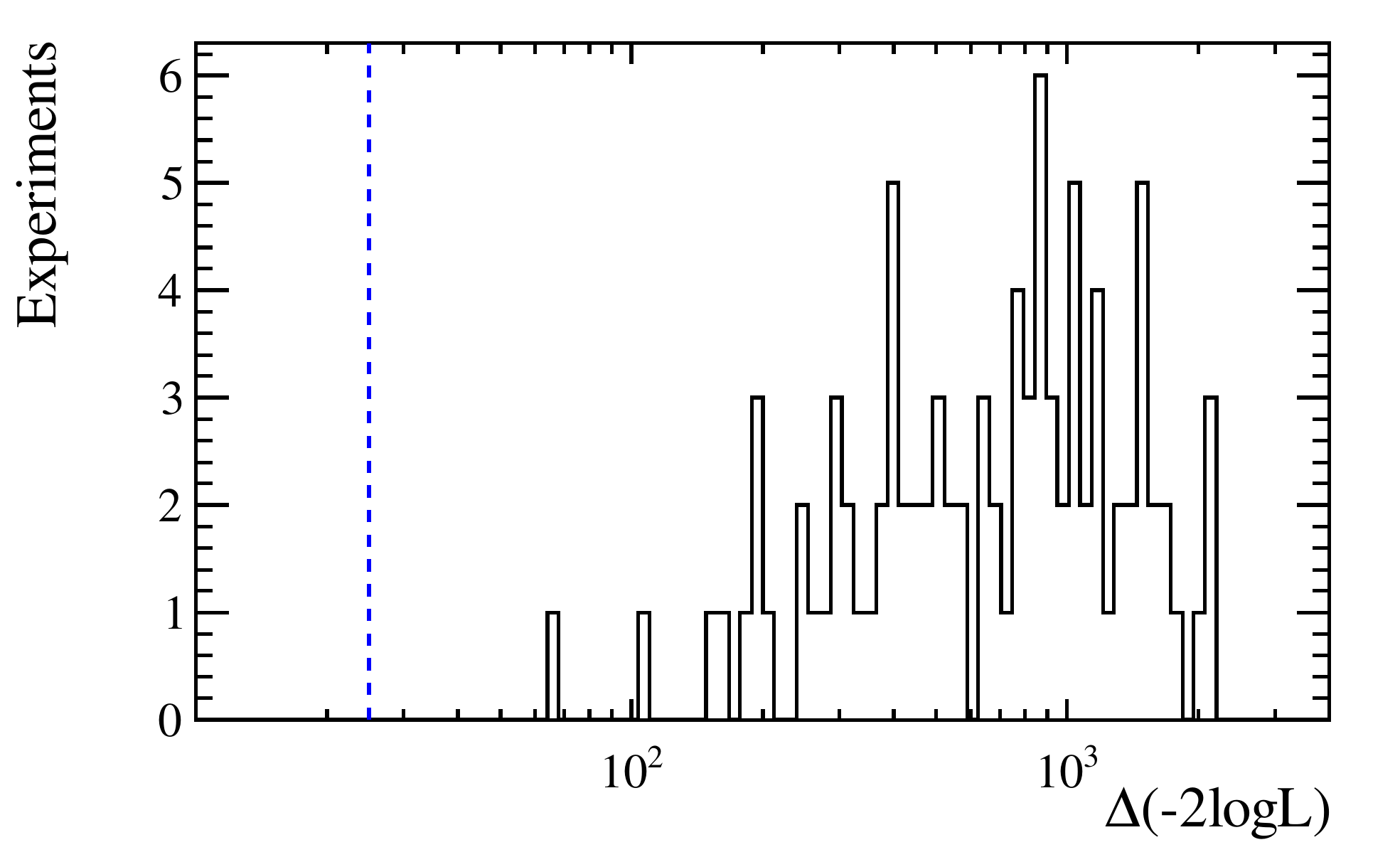}
  \put(-230,100){(c)}
  \put(-32,100){(d)}

  \caption{\label{fig:dll} Difference in $-2\log{\cal L}$ between the two solutions for $\phitwo^{00}$ expected with (a) Run 2 data at LHCb (b) 10 ab$^{-1}$ at Belle II (c) Run 3 data at LHCb (d) 50 ab$^{-1}$ at Belle II. The vertical dashed line indicates the cut-off at which a $5\sigma$ distinction is achieved. Note the log scale on the bottom plots.}
\end{figure}

I then perform two fits to each pseudo-experiment. The first sets the starting value of $\phitwo^{00}$ to its generated value, the second sets it to the second solution. The distribution of the difference in $-2\log{\cal L}$ is then recorded in figure~\ref{fig:dll}. The spread of these distributions is a measure of the statistical and specifically considered systematic effects at play in this study, namely those arising from the uncertainties of the parameters listed in tables~\ref{tab:stage1} and \ref{tab:stage2} that are varied in generation.

In the case of data recorded at LHCb up to the end of Run 2, a $2\sigma$ separation between $\phitwo^{00}$ solutions appears to be the most likely outcome. Corresponding to a scenario similar to that shown in figure~\ref{fig:phi2_err}c, this implies that a \phitwo\ solution could be ruled out at the $1\sigma$ level in $B \to \rho\rho$, which would already mark quite an important achievement. Due to the asymmetric nature of the log-likelihood difference, there is still over a 50\% chance that the $\phitwo^{00}$ solutions could be resolved even at the $5\sigma$ level, at which point it will become important to improve measurements involving the other $B \to \rho \rho$ channels in order to further suppress multiple solutions for \phitwo. Under the assumptions laid out in this study, a $5\sigma$ separation between solutions would most likely be borderline with a data sample of 10 ab$^{-1}$ collected by Belle II, while 50 ab$^{-1}$ or Run 3 data from LHCb, should be more than adequate.

\begin{figure}[tbp]
  \centering
  \includegraphics[height=120pt,width=!]{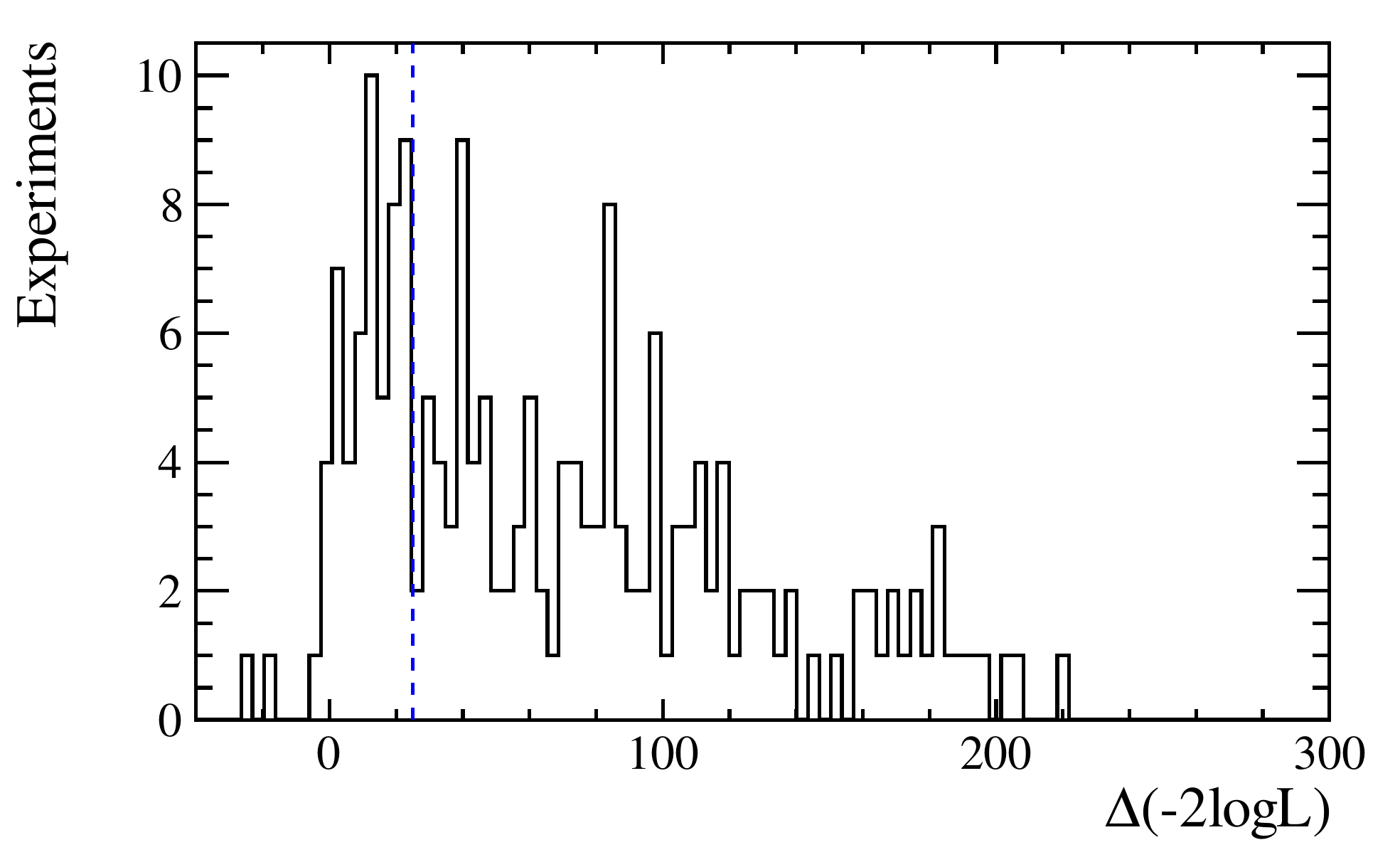}
  \includegraphics[height=120pt,width=!]{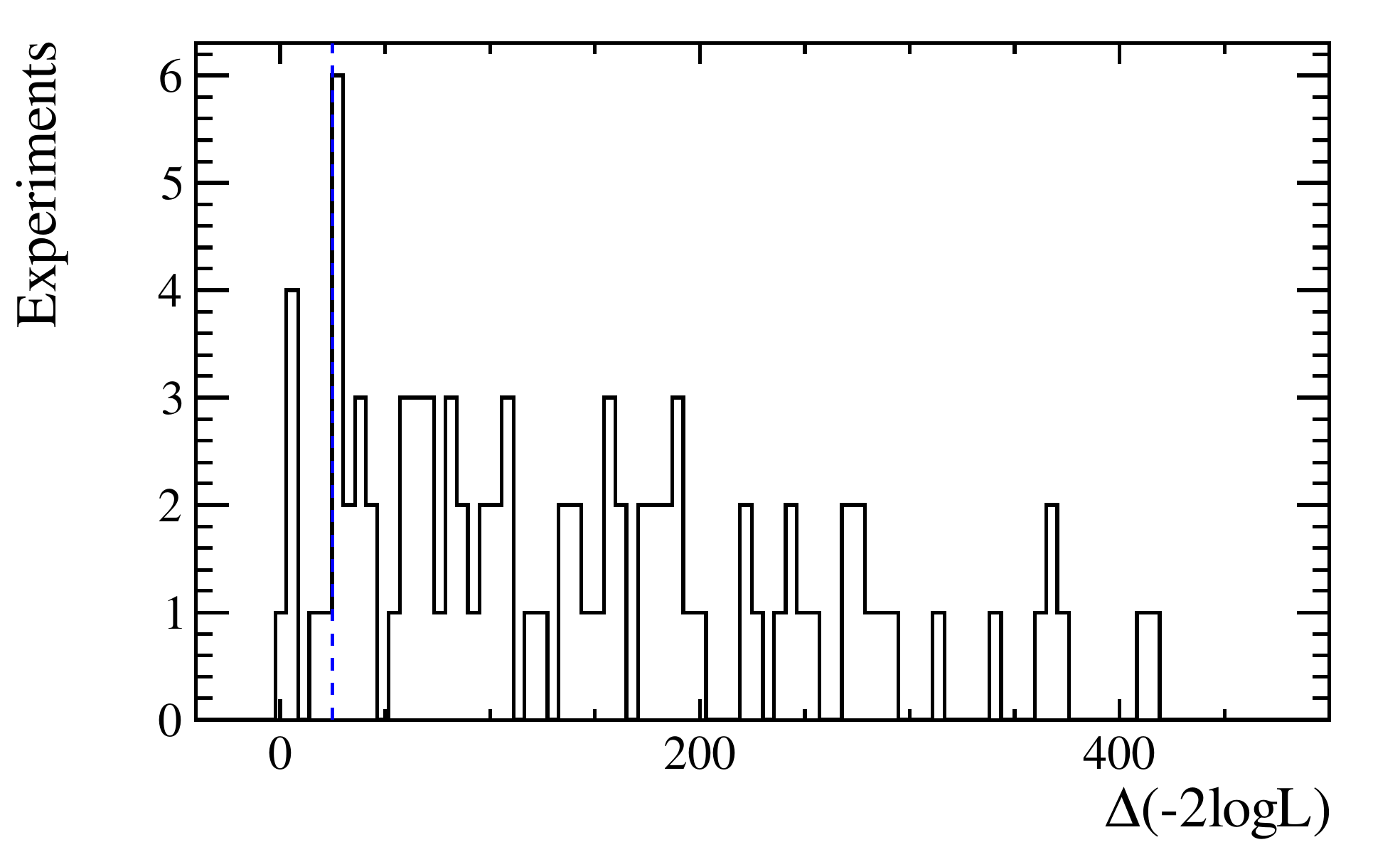}
  \put(-230,100){(a)}
  \put(-32,100){(b)}

  \includegraphics[height=120pt,width=!]{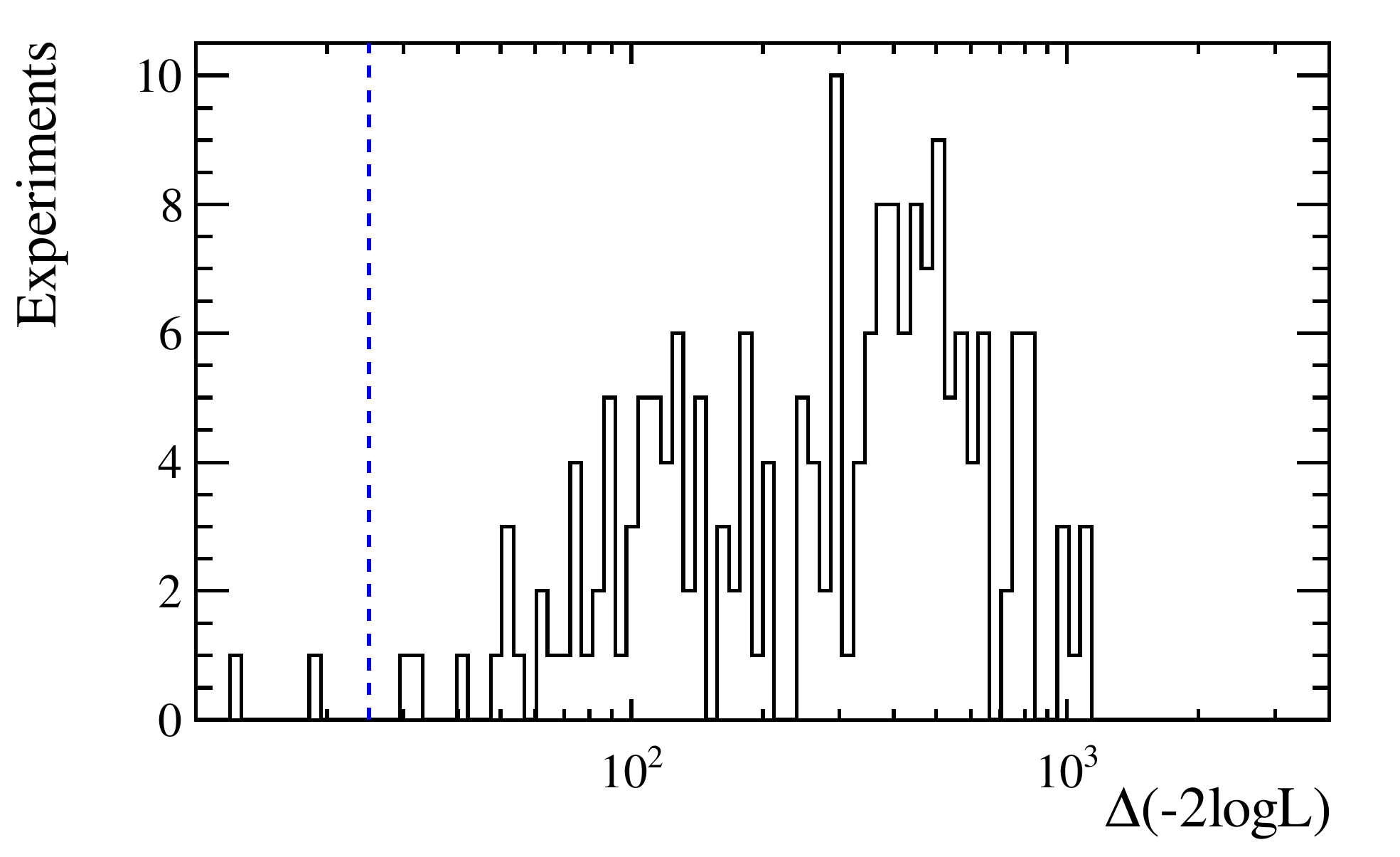}
  \includegraphics[height=120pt,width=!]{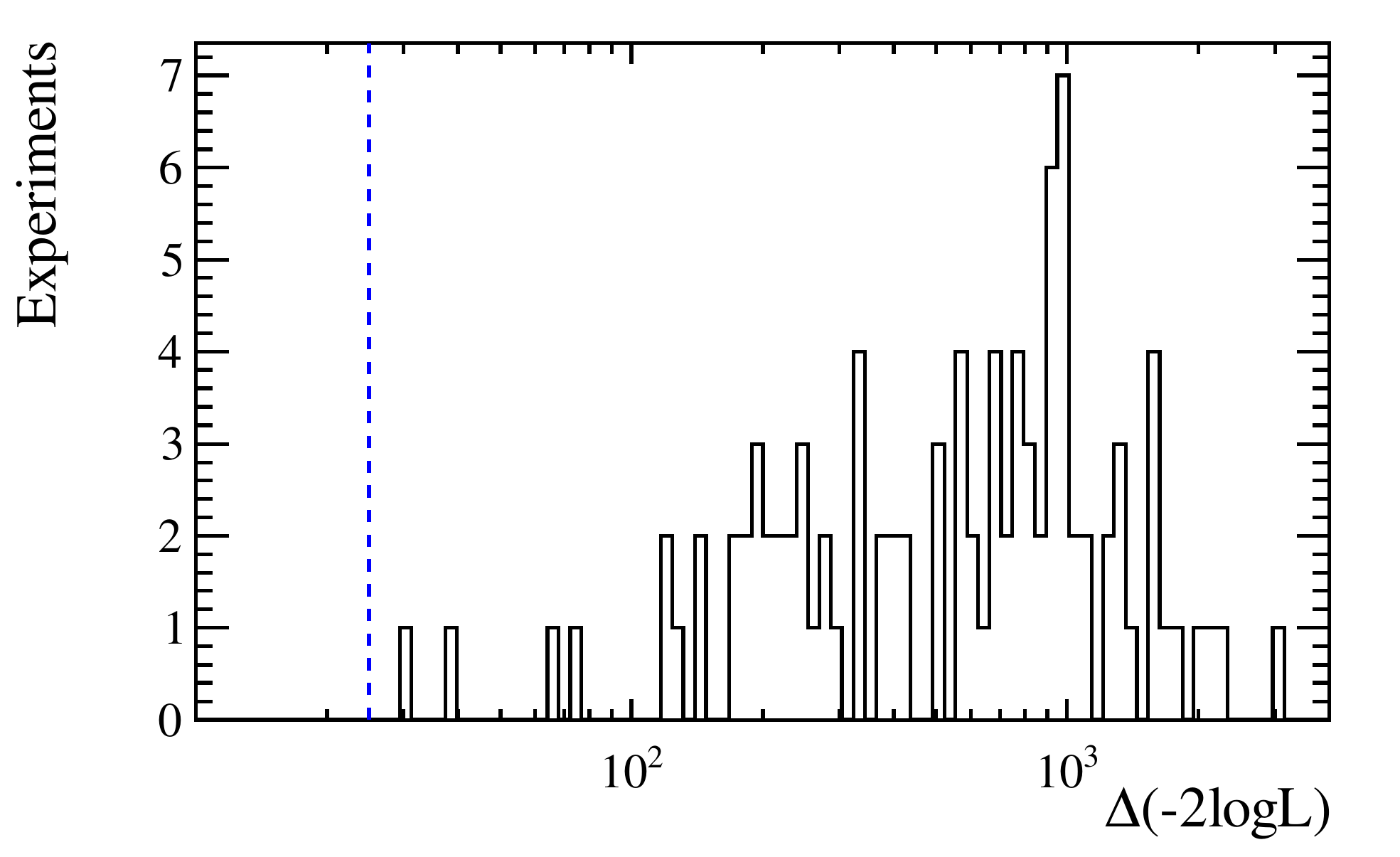}
  \put(-230,100){(c)}
  \put(-32,100){(d)}

  \includegraphics[height=120pt,width=!]{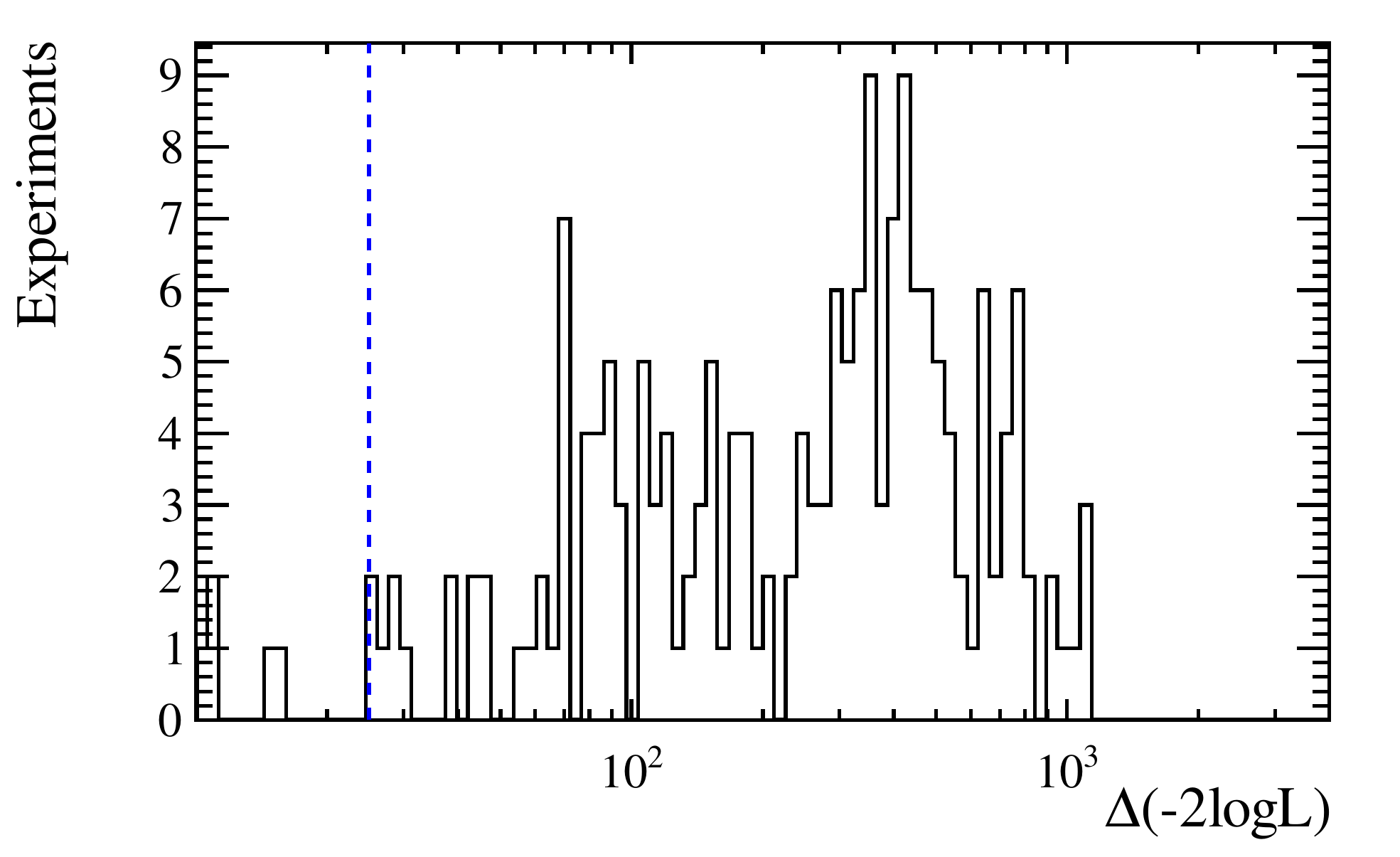}
  \includegraphics[height=120pt,width=!]{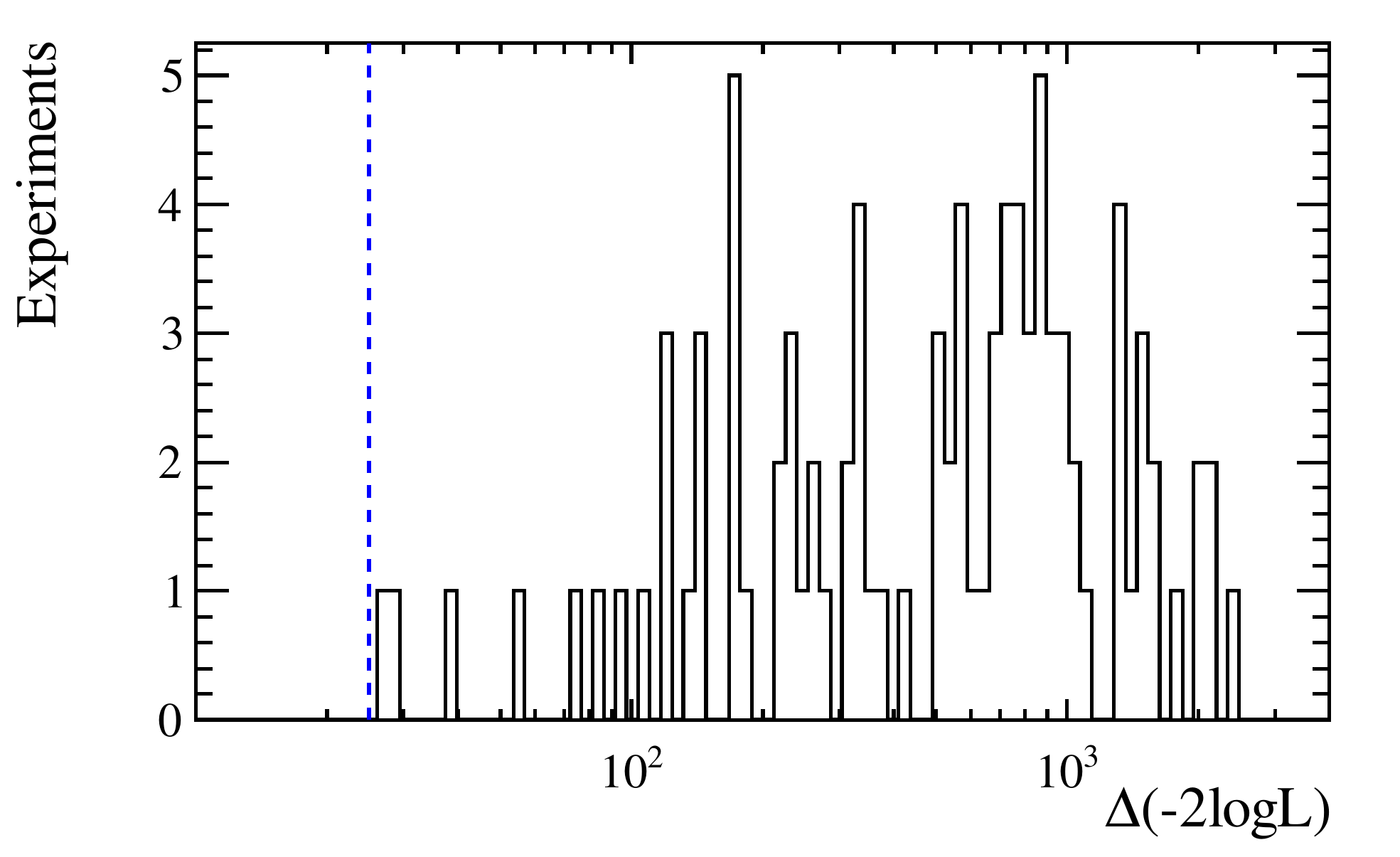}
  \put(-230,100){(e)}
  \put(-32,100){(f)}

  \caption{\label{fig:dllalt} Difference in $-2\log{\cal L}$ between the two solutions for $\phitwo^{00}$ expected with Run 3 data at LHCb shown on the left and 50 ab$^{-1}$ at Belle II on the right. (a) and (b) represent the test with the LHCb $\aone$ cut, (c) and (d) the case of no $CP$ violation in $\Bz \to \rhoz \rhoz$, while in (e) and (f), $\phitwo^{00}$ is set to the $\Bz \to a_1^+ \pim$ weak phase. The vertical dashed line indicates the cut-off at which a $5\sigma$ distinction is achieved. Note the log scale on the bottom four plots.}
\end{figure}

In the remaining, I consider additional scenarios that may be of interest. To highlight the impact of the selection imposed to suppress the \aone\ contribution in the LHCb analysis, I repeat the study with their criteria on the 3-pion mass. Figure~\ref{fig:dllalt}a,b indicates that this would indeed constitute a risky venture at LHCb even at the end of Run 3, while at Belle II a clear separation of $\phitwo^{00}$ solutions is by no means assured with the full data set.

I also examine the situation when no $CP$ violation is present in $\Bz \to \rhoz \rhoz$, with $|\lambda^{00}| = 1$ and $\phitwo^{00} = 88.8^\circ$, taken from the CKMfitter average for the direct measurement of $\phitwo$~\cite{CKMFitter}. Compared with the $CP$ violating case shown in figure~\ref{fig:dll}c,d, the absence of $CP$ violation doesn't have an immediately obvious effect as portrayed in figure~\ref{fig:dllalt}c,d. Finally, I track the difference in log-likelihood when $\phitwo^{00}$ exactly aligns with the generated weak phase of the dominant $\Bz \to a_1^+ \pim$ contribution. Again, no appreciable difference apart from perhaps a minor degradation in performance can be seen in figure~\ref{fig:dllalt}e,f against the nominal case on display in figure~\ref{fig:dll}c,d. These additional tests suggest that the ability to resolve $\phitwo^{00}$ solutions rests more on the presence of the $\Bz \to \aone \pimp$ amplitudes and their $CP$ violation parameters rather than the $CP$ violation parameters of $\Bz \to \rhoz \rhoz$ itself.

In future analyses, further intermediate states beyond those considered here are expected to materialise as data samples increase in size. At the very least, they could have no effect on the ability to resolve the $\Bz \to \rhoz \rhoz$ solution degeneracy in the unlikely case that the effective weak phase of each additional contribution aligns exactly with either $\phitwo^{00}$ or the weak phases of $\Bz \to a_1^\pm \pimp$. In reality however, it is more likely that each will carry a unique effective weak phase depending on the strength of their respective penguin couplings, which can only improve the odds of resolving a single solution in $\Bz \to \rhoz \rhoz$ over what is demonstrated in this study.

\section{Conclusion}
\label{sec:conc}
I present an extension to the SU(2) isospin triangle analysis, that has the capacity to resolve the \phitwo\ degeneracy in the $B \to \rho\rho$ system under certain conditions. For this purpose a time-dependent flavour-tagged amplitude analysis involving $\Bz \to \rhoz\rhoz$ decays is proposed, where the phase space must be expanded over that used in current analyses to allow its interference with $\Bz \to \aone \pimp$ to be sufficiently understood. If meaningful discrimination between the two solutions of the effective $\phitwo$ in $\Bz \to \rhoz\rhoz$ can be achieved and its central value is not located in the middle of the two solutions for the effective $\phitwo$ coming from $\Bz \to \rho^+ \rho^-$, then the \phitwo\ solution degeneracy in $B \to \rho\rho$ can be resolved.

The current experimental uncertainty in $\Bz \to \aone \pimp$ combined with a theoretical prediction of its own effective \phitwo, indicates that the benefits of potential deviation from tree-level expectations as would be induced by penguin contributions to these channels, are sufficient to warrant the additional experimental complications involved. Under varying assumptions on the $CP$ violation present in $\Bz \to \rhoz\rhoz$, the main hadronic uncertainties due to the \aone\ also appear to be under control as studies with roughly estimated signal yields at LHCb and Belle II indicate a real possibility to resolve \phitwo\ in the $B \to \rho\rho$ system at the end of Run 2 and with 10 ab$^{-1}$, respectively.

\acknowledgments{
  I am grateful to P.~Vanhoefer for our discussions arising from his work on $\Bz \to \rhoz \rhoz$ while I studied $\Bz \to \aone \pimp$ at Belle, which inspired this idea. Thanks also goes out to J.~Rademacker, whose pioneering methodology developed for 4-body Dalitz Plot analyses provided me with the knowledge necessary to evaluate the feasibility of this method. I am indebted to D.~Mart\'{i}nez~Santos and P.~Naik for commenting on the paper draft and also to J.~Albrecht who checked the projections on behalf of the LHCb physics coordination. Finally, special thanks to T.~Gershon, whose careful reading improved this work immensely. This work was supported by XuntaGal (Spain).
}

\end{document}